\documentclass[10pt,journal,compsoc]{IEEEtran}

\ifCLASSOPTIONcompsoc
  \usepackage[nocompress]{cite}
\else
  \usepackage{cite}
\fi

\usepackage{amsmath}
\usepackage{xspace}
\usepackage{subfig}
\usepackage{multirow}
\usepackage{enumerate}
\usepackage{enumitem}
\usepackage{color}
\usepackage{graphicx}
\usepackage{wasysym}
\usepackage{balance}
\usepackage{bm}
\usepackage{url}
\usepackage{booktabs}
\usepackage{amssymb}

\newtheorem{definition}{Definition}

\usepackage{threeparttable}
\usepackage{dcolumn}
\newcolumntype{d}[1]{D{.}{.}{#1}}

\newcommand{\eat}[1]{}
\newcommand{\paratitle}[1]{\vspace{1ex}\noindent \textbf{#1}}

\let\oldhat\hat
\renewcommand{\vec}[1]{\textbf{#1}}
\renewcommand{\hat}[1]{\oldhat{\textbf{#1}}}
\renewcommand{\matrix}[1]{\textbf{#1}}

\newcommand{\eg}{\emph{e.g.,}\xspace}
\newcommand{\rf}{\emph{rf.}\xspace}

\newcommand{\ie}{\emph{i.e.,}\xspace}

\newcommand{\etal}{\emph{et al.}\xspace}

\begin{document}
\title{Exploring Global Information for Session-based Recommendation}

\author{Ziyang~Wang,
        Wei~Wei$^\dagger$,
        Gao~Cong,
        Xiao-Li~Li,
        Xian-Ling~Mao,
        Minghui~Qiu
\thanks{This paper is an extended version of the SIGIR '20 conference paper~\cite{wang2020global}.
Wei~Wei is the corresponding author.}
\IEEEcompsocitemizethanks{
\IEEEcompsocthanksitem Ziyang~Wang and Wei~Wei was with Cognitive Computing and Intelligent Information Processing (CCIIP) Laboratory, School of Computer Science and Technology,Huazhong University of Science and Technology, Wuhan, China. E-mail: \{ziyang1997,weiw\}@hust.edu.cn.
\IEEEcompsocthanksitem Gao~Cong was with School of Computer and Engineering, Nanyang Technological University, Singapore. E-mail: gaocong@ntu.edu.sg.
\IEEEcompsocthanksitem Xiao-Li Li was with Institute for Infocomm Research, Singapore. E-mail: xlli@i2r.a-star.edu.sg.
\IEEEcompsocthanksitem Xian-Ling~Mao was with School of Computer Science and Technology, Beijing Institute of Technology, Beijing, China. E-mail: maoxl@bit.edu.cn.
\IEEEcompsocthanksitem Minghui Qiu was with Alibaba Group, Hangzhou, China.
E-mail: minghuiqiu@gmail.com.
}
}

\markboth{Journal of \LaTeX\ Class Files,~Vol.~14, No.~8, August~2015}%
{Shell \MakeLowercase{\textit{et al.}}: Bare Demo of IEEEtran.cls for Computer Society Journals}
%



\IEEEtitleabstractindextext{%
\begin{abstract}
Session-based recommendation (SBR) is a challenging task,
which aims at recommending items based on anonymous behavior sequences.
Most existing SBR studies model the user preferences based only on the current session while neglecting the item-transition information from the other sessions, which suffer from the inability of modeling the complicated item-transition pattern. To address the limitations, we introduce global item-transition information to strength the modeling of the dynamic item-transition.
For fully exploiting the global item-transition information, two ways of exploring global information for SBR are studied in this work.
Specifically, we first propose a basic GNN-based framework~(BGNN), which solely uses session-level item-transition information on 
\emph{session} graph.
Based on BGNN, we propose a novel approach, called \textbf{S}ession-based \textbf{R}ecommendation with \textbf{G}lobal \textbf{I}nformation (\textsf{SRGI}), which
infers the user preferences via fully exploring global item-transitions over all sessions from two different perspectives:
(i) Fusion-based Model~\textsf{(SRGI-FM)}, which recursively incorporates the neighbor embeddings of each node on \emph{global} graph into the learning process of session-level item representation;
and
(ii) Constrained-based Model~\textsf{(SRGI-CM)}, which treats the global-level item-transition information as a constraint to ensure the learned item embeddings are consistent with the
global item-transition.
Extensive experiments conducted on three popular benchmark datasets
demonstrate that both \textsf{SRGI-FM} and \textsf{SRGI-CM} outperform the state-of-the-art methods consistently.
\end{abstract}

}

\maketitle

\IEEEdisplaynontitleabstractindextext

\IEEEpeerreviewmaketitle

\section{Introduction}

With the explosion of information on the Internet, recommendation systems play an increasingly important role as they recommend useful content to address the information overload problem.
Conventional recommendation approaches (\eg collaborative filtering\cite{mnih2008probabilistic,kabbur2013fism,hsieh2017collaborative}) usually make predictions based on the user profiles and long-term interaction history. However, in many recent real-world scenarios (\eg on-line shopping platform\footnote{\url{https://www.amazon.com/}} and mobile stream media\footnote{\url{https://www.tiktok.com/}}), such information may not exist.
In addition, users’ recent behaviors indicate their current interests, which is neglected by these methods.
To bridge the gaps in conventional recommendation approaches, session-based recommendation~(SBR) is emerged to predict the next action (\eg click) of users based on anonymous behavior sequences in chronological order.

Due to its highly practical value, SBR has attracted increasing attention and many approaches have been proposed.
Most of early studies on session-based recommendation  
are based on Markov-chains\cite{rendle2010factorizing}, which predict the users' next interest solely based on the previous
action. However, these methods rely on strong independence assumption, which may suffer from noisy data and intractable computation problem for real-world applications.

Recent years have witnessed the rapid development of deep learning techniques~\cite{he2017neural,liang2018variational,li2020learning} and many neural networks based approaches are proposed for SBR, which model the item-transition within the current session and can be roughly categorized into two types, \ie RNN-based and GNN-based. The former~\cite{hidasi2015session,li2015gated,ijcai2019-523} regard SBR as a sequence modeling task and leverage recurrent neural networks (RNN) to capture the item-transition information in chronological order, which is then extended with attention network~\cite{li2017neural} and memory network~\cite{liu2018stamp}. The latter~\cite{wu2019session,qiu2019rethinking,yu2020tagnn} focus on capturing the dependencies between each item and its context (\ie adjacent items) in the session, which first convert each session into a subgraph and then employ graph neural network (\eg GGNN~\cite{li2015gated}, GAT~\cite{velivckovic2017graph}) to extract the item features from the graph. 

\begin{figure}[!t]
  \centering
  \includegraphics[width=\columnwidth, angle=0]{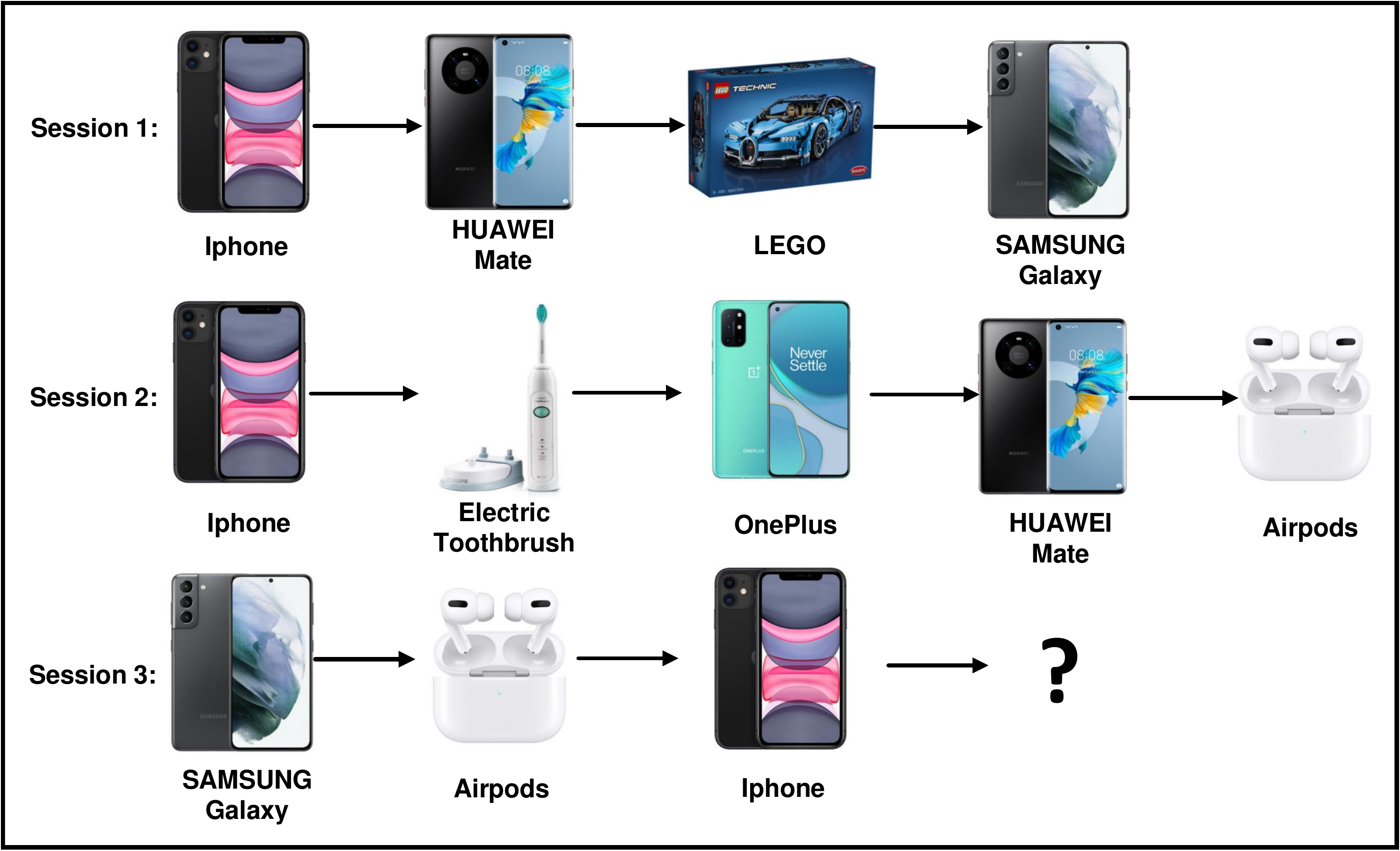}
  \caption{Example of item-transition of sessions.}
  \label{fig:frameWorkexample}
\end{figure}

Although encouraging results are achieved, these methods still suffer from the inability of modeling the complicated inherent order of item-transition for each session, due to the noisy data and limited session information. Recently, some efforts~\cite{wang2019collaborative,ijcai2020-359} are devoted to incorporating collaborative information into SBR to strength the modeling of item-transition. 
Specifically, CSRM~\cite{wang2019collaborative} fuses the latest $m$ sessions to enrich the representation of the current session via memory network. CoSAN~\cite{ijcai2020-359} injects feature embedding represented by neighborhood sessions to enrich the item representation in the current session. 
However, they may unfortunately encode both relevant and irrelevant information of the other sessions into the current session representation, which may even deteriorate the performance~\cite{ijcai2019-523}.
To illustrate this, an example is shown in Figure~\ref{fig:frameWorkexample}.
Without loss of generality,
suppose the current session is ``Session 3'', and the session-based recommendation aims to recommend the next interest of the user based on the past three clicked items.
From Figure~\ref{fig:frameWorkexample}, we observe that:
(i) 
Utilizing the item-transition of the other sessions might help model
the user preference of the current session.
For example, from ``Session 1'' and ``Session 2'' we can find relevant items from item-transition information for ``Iphone'', such as ``HUAWEI Mate'' and ``OnePlus''.
and (ii) Directly utilizing the item-transition information of the \emph{entire} other session may introduce noise
when part of the item-transition information encoded in such session is not relevant to the current session.
For instance,
CSRM~\cite{wang2019collaborative} and CoSAN~\cite{ijcai2020-359} may also consider to utilize ``Session 1'' and ``Session 2'' to help modeling the user preference of ``Session 3'',
which will introduce the \emph{irrelevant} items (\ie ``LEGO'' and ``Electric Toothbrush'')
when learning ``Session 2'''s embedding as it treats other session as a whole without distinguishing relevant item-transition from irrelevant item-transition, which is challenging.

To this end, we propose a novel approach to exploit the item-transitions over all sessions from two perspectives
for better inferring the user preference of the current session for session-based recommendation, which is named \textbf{S}ession-based \textbf{R}ecommendation with \textbf{G}lobal \textbf{I}nformation (\textsf{SRGI}).
%
%
In \textsf{SRGI}, we propose to learn two levels of item feature from \emph{session} graph and \emph{global} graph, respectively:
(i) \emph{Session graph}, which is to learn the session-level item feature by modeling pairwise item-transitions within the current session;
and (ii) \emph{Global graph}, which is to learn the global-level item feature by modeling pairwise item-transitions over all sessions~(including the current session).
%
\textsf{SRGI} first employs a basic GNN framework~(B-GNN) on the \emph{session} graph to learn \emph{session}-level item embedding within the current session from three kinds of relations.
Then based on the constructed \emph{global} graph, we present two versions of SRGI that incorporates the global transition information into SBR:
(i) Fusion-based Model~\textsf{(SRGI-FM)}, which directly incorporates the neighbors' features of each node on the global graph through a session-aware attention mechanism to enrich the features of current session;
and (ii) Constraint-based Model~\textsf{(SRGI-CM)}, which preserves the global proximity to ensure the learnt item embeddings are consistent with the global item-transition,
via graph contrastive learning.

The main contributions of this work are summarized as follows:

\begin{itemize}
\item We propose a basic GNN-based session recommendation framework for SBR, which employs graph neural networks layer to explicitly extracting item features derived from three kinds of relations on session-level graph.

\item We propose a novel unified model (named \textsf{SRGI}) to fully and effectively explore  the pairwise item-transition information from two levels of graph models, \ie \emph{session} graph and \emph{global} graph from two different aspects,
    \ie \emph{Fusion-based Model} that circularly incorporates the global neighbor information into the item embedding learning process for enhancing the session-level item representations,
   and \emph{Constraint-based Model} treats the global-level item-transition information as a constraint and employ graph contrastive learning to ensure the learnt item embeddings are consistent with the global graph structure;

\item We conduct extensive experiments on three real-world datasets, which demonstrate the efficacy of \textsf{SRGI-FM} and \textsf{SRGI-CM} over state-of-the-art baselines.
\end{itemize}

\section{Related Work}

\subsection{Session-based Recommendation}

Many research efforts have been conducted on session-based recommendation, which will be reviewed in this section.

\paratitle{Markov Chain-based SBR}.
Several traditional methods can be employed for SBR although they are not originally designed for SBR. 
For example, Markov Chain-based methods map each session into a Markov chain, where the user's next action is inferred based on the previous one. 
Shani \etal \cite{shani2005mdp} utilize Markov Decision Processes (MDP) to model the transitions of items within a session, where simplest MDP boils down to first-order Markov chains \cite{ren2019repeatnet}.
Rendle \etal \cite{rendle2010factorizing} propose FPMC to capture both sequential patterns and long-term user preference
by a hybrid method based on the combination of matrix factorization and first-order Markov chain for recommendation. It can be adapted for SBR by ignoring the user latent representation as it is not available for anonymous SBR.
However, MC-based methods usually focus on modeling sequential transition of two adjacent  items.
In contrast, our proposed model converts the sequentially item-transitions into graph-structure data for capturing the inherent order of item-transition patterns for SBR.

\paratitle{Deep-learning based SBR}.
In recent years, neural network-based methods that are capable of  modeling sequential data have been utilized for SBR.
Hidasi \etal~\cite{hidasi2015session} propose the first work called GRU4REC
to apply the RNN networks for SBR, which adopts a multi-layer Gated Recurrent Unit (GRU) to model item interaction sequences.
Then, Tan \etal \cite{tan2016improved} extend the method \cite{hidasi2015session} by introducing data augmentation.
Li \etal \cite{li2017neural} propose NARM that
incorporates attention mechanism into stack GRU encoder to capture the more representative item-transition information for SBR.
Liu~\etal~\cite{liu2018stamp} propose an attention-based short-term memory networks (named STAMP)
to captures the user's current interest without using RNN. Both NARM and STAMP emphasize the importance of the last click by using attention mechanism.
Inspired by $Transformer$ \cite{vaswani2017attention}, SASRec \cite{kang2018self} stacks multiple layers to capture the relevance between items.
ISLF \cite{ijcai2019-799} takes into account the user's interest shift, and employs  variational auto-encoder (VAE) \cite{pu2016variational} and RNN
to capture the user's sequential behavior characteristics for SBR.
MCPRN \cite{ijcai2019-523} proposes to model the multi-purpose of a given session by using a mixture-channel model for SBR.
%
%
However, similar to MC-based methods, RNN-based methods focus on modeling the sequential transitions of adjacent items~\cite{ijcai2019-883}
to infer user preference via the chronology of the given sequence, and thus cannot model the complex item-transition patterns (e.g., non-adjacent item transitions).

Recently, several proposals employ GNN-based model on graph built from the current session
to learn item embeddings for SBR.
Wu~\etal~\cite{wu2019session} convert each session into a graph and leverage gated GNN to the explore the complex transitions among items.
Following the success of SR-GNN, some variants are also proposed for SBR,
such as GC-SAN \cite{ijcai2019-547}, which combine GNN layers and self attention layers to learn the dependencies within the session.
%
Qiu~\etal \cite{qiu2019rethinking} propose FGNN to learn each item representation by aggregating its neighbors' embeddings with multi-head attention.
Yu~\etal \cite{yu2020tagnn} leverage target-aware attention to dynamically obtain the importance of each item within the session.
Meng~\etal \cite{meng2020incorporating} incorporate knowledge graph to capture the dependencies between items.
However, all these approaches only model the item-transition information on the current session.
In contrast, our proposed model learns the item-transition information over all sessions to enhance the modeling of item-transition and inference of user interests.

\paratitle{Collaborative Filtering-based SBR}.
Although deep learning based methods have achieved remarkable performance,
collaborative filtering (CF) based methods can still provide competitive results.
Item-KNN \cite{sarwar2001item} can be extended for SBR by recommending items that are most similar to the last item of the current session.
%
KNN-RNN~\cite{jannach2017recurrent} makes use of GRU4REC~\cite{hidasi2015session}
and the co-occurrence-based KNN model to extract the sequential patterns for SBR.
CSRM~\cite{wang2019collaborative} first utilizes NARM over item-transitions to encode each session, then enriches the representation of the current session by exploring the latest $m$ neighborhood sessions.
CoSAN~\cite{ijcai2020-359} incorporates neighborhood sessions embedding into the items of current session and employ multi-head self-attention to capture the dependencies between each item.
However, CSRM and CoSAN may suffer from noise when integrating other sessions' embeddings for the current one.
In contrast,  our proposed method considers the collaborative information in \emph{item-level}: we use the item embeddings in other sessions to enrich the item embeddings of the current session,
and then integrate them into session representation for SBR.

\subsection{Graph Contrastive Learning}

Graph contrastive learning aims to learn the discriminative node representations by contrasting positive and negative samples in graph. 
Early work on graph contrastive learning focus on capturing local structural patterns, which forces neighboring nodes to have similar feature representations.
For example, DeepWalk\cite{perozzi2014deepwalk} and node2vec\cite{grover2016node2vec} obtain walk sequences on the graph and consider nodes within the same window in the sequence as positive samples.
However, random-walk based approaches overly emphasize the structural information \cite{qiu2018network} and their performance are heavily affected by hyperparameter choice.

Recently, with the development of deep learning technique, some studies employ graph neural networks to obtain better item representations for graph constrastive learning.
Based on the powerful graph convolutional architectures, 
DGI \cite{velickovic2019deep} maximizes the mutual information between global and local representations. 
GMI \cite{peng2020graph} focus on the mutual information between input and representation of both the node and edge in the graph. 
Via data augmentations, GraphCL \cite{you2020graph} maximizes the agreement between two augmented views of the input graph.
Considering adaptive graph augmentations, GCA \cite{zhu2020graph} incorporates priors for topological and semantic aspects of the graph.

\section{Preliminaries}
In this section, we first present the problem statement,
and then introduce two types of graph, \ie \emph{session} graph
and \emph{global} graph,
based on different levels of pair-wise item transitions over sessions
for learning item representations,
in which we highlight the modeling of \emph{global}-level item transition
information as it is the basis of \emph{global} graph construction.
For clarity, frequently used notations are summarized in Table \ref{tbl-symbol-and-definitionn}.

\begin{table}[!t]
\centering
\caption{Notations and Definitions.}
\footnotesize\label{tbl-symbol-and-definitionn}{
\begin{tabular}{c||p{170pt}} \hline
\textbf{Notation} & \textbf{Definition}   \\ \hline \hline
  \textbf{Input} & \\ \hline
        $S$ & Anonymous session, $S = \{ v^{s}_{1}, v^{s}_{2}, ..., v^{s}_{l} \}$ \\ \hline
        \textbf{Graph} & \\ \hline
        $ \mathcal{G}_{s}, \mathcal{G}_{g} $ & Session graph and global graph \\
        $ \mathcal{V}_{s}, \mathcal{V}_{g} $ & Nodes in session graph and global graph \\
        $ \mathcal{E}_{s}, \mathcal{E}_{g} $ & Edges in session graph and global graph \\
        $ \mathcal{N}^s_v $ & Neighbor set of node $v$ in session graph \\
        $ \mathcal{N}_{\varepsilon}({v}) $ &  $\varepsilon-$neighbor set of node $v$ in global graph \\


        \hline \textbf{Latent Variable} & \\ \hline
        $ \vec{h}_{\mathcal{N}^g_{v_i}} $ & Features of the neighbors of the item $v_i$ obtained from global graph \\
        $ \vec{h}^{g}_{v_i} $ & Features of item $v_i$ obtained from global graph \\
        $ \vec{h}^s_{v_i} $ & Features of item $v_i$ obtained from session graph \\
        $ \vec{h}^{\prime}_{v_i} $ & Final representation of the item $v^s_i$ \\
        $ \vec{S} $ & Final representation of the current session \\
        $ \vec{o} $ & Features of items from graph views after data augmentation \\
        \hline \textbf{Output} & \\ \hline
        $ \hat{\vec{y}} $ & Output probabilities of each item\\
        
        \hline \textbf{Loss Function} & \\ \hline
        $ L_p $ & Cross-entropy of the prediction results\\
        $ L_c $ & Contrastive loss for the global graph structure \\
        \hline
\end{tabular}}
\end{table}
\subsection{Problem Statement}


%
Let  $V= \{ v_{1}, v_{2}, ..., v_{m}\}$ be all of items,
and each session S be denoted by
%
$S = \{ v^{s}_{1}, v^{s}_{2}, ..., v^{s}_{l}\}$, consisting of a sequence of interactions (\ie items clicked by a user) in chronological order,
where $v^{s}_{i}$ denotes an item clicked at time-step $i$ within session $S$, and the length of $S$ is $l$.
Given a session $S$, the problem of session-based recommendation aims to
recommend the top-$N$ items ($1 \leq N \leq |V|$)
from $V$ that are most likely to be clicked by the user in the next timestamp~(\ie $v^{s}_{l+1}$).

\subsection{Graph Models: Session Graph and Global Graph}
In this subsection, we present two different graph models to capture different levels of \emph{item transition} information
over all available sessions for item representation learning.

\subsubsection{Session Graph Model}
Session-based graph aims to learn the \textbf{\emph{session-level}} item embedding
by modeling sequential patterns over pair-wise adjacent items in the current session.
Inspired by \cite{wu2019session}, each session sequence
is converted into a session graph for learning
GNN-based item embeddings of the current session.
%
%
More concretely,
for each session $S = \{ v^{s}_i\}^{l}_{i=1}$,
the corresponding \emph{\textbf{session graph}} is defined as a 2-tuple
\begin{math}
  \mathcal{G}_{s} = ( \mathcal{V}_{s}, \mathcal{E}_{s} ),
\end{math}
where $\mathcal{V}_{s}\subseteq V$ and $\mathcal{E}_{s}=\{e^{s}_{ij}\}$ are denoted the clicked item set and the edge set in $S$ respectively, and
\begin{math}
    e^{s}_{ij} = (v^{s}_i, v^{s}_j)
\end{math}
indcates the adjacent edge of node $v^{s}_i$ and $v^{s}_j$ in $S$,
which is called \textbf{\emph{session}-level} item-transition pattern.

\begin{figure}[!t]
 \subfloat[\textbf{\emph{Session}} Graph.]{
  \begin{minipage}[b]{\linewidth}
  \centering
  \includegraphics[width=\linewidth]{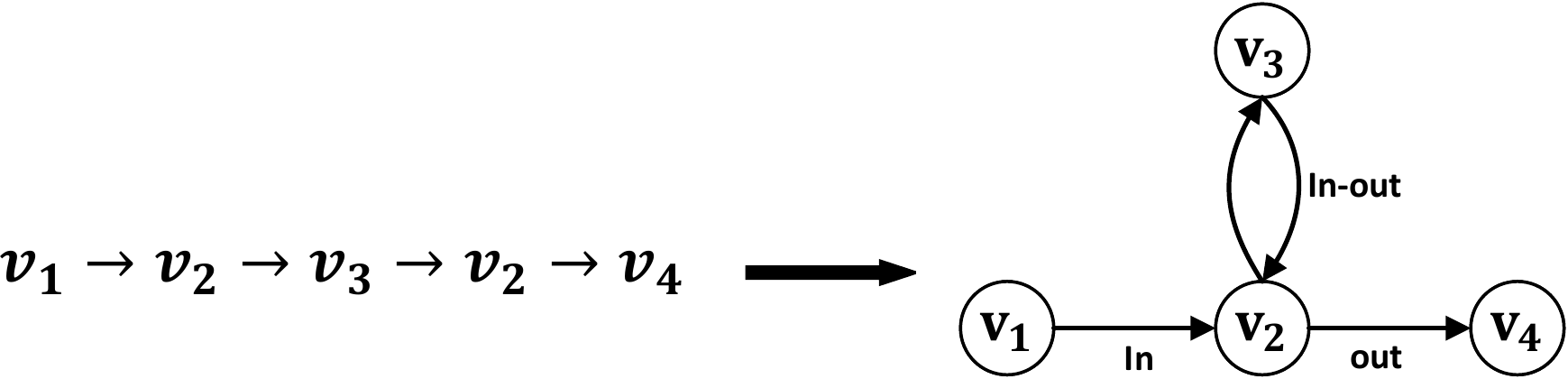}
  \end{minipage}
  \label{sessionGraph}
 }

 \subfloat[\textbf{\emph{Global Graph}}.]{
  \begin{minipage}[b]{\linewidth}
  \centering
  \includegraphics[width=\linewidth]{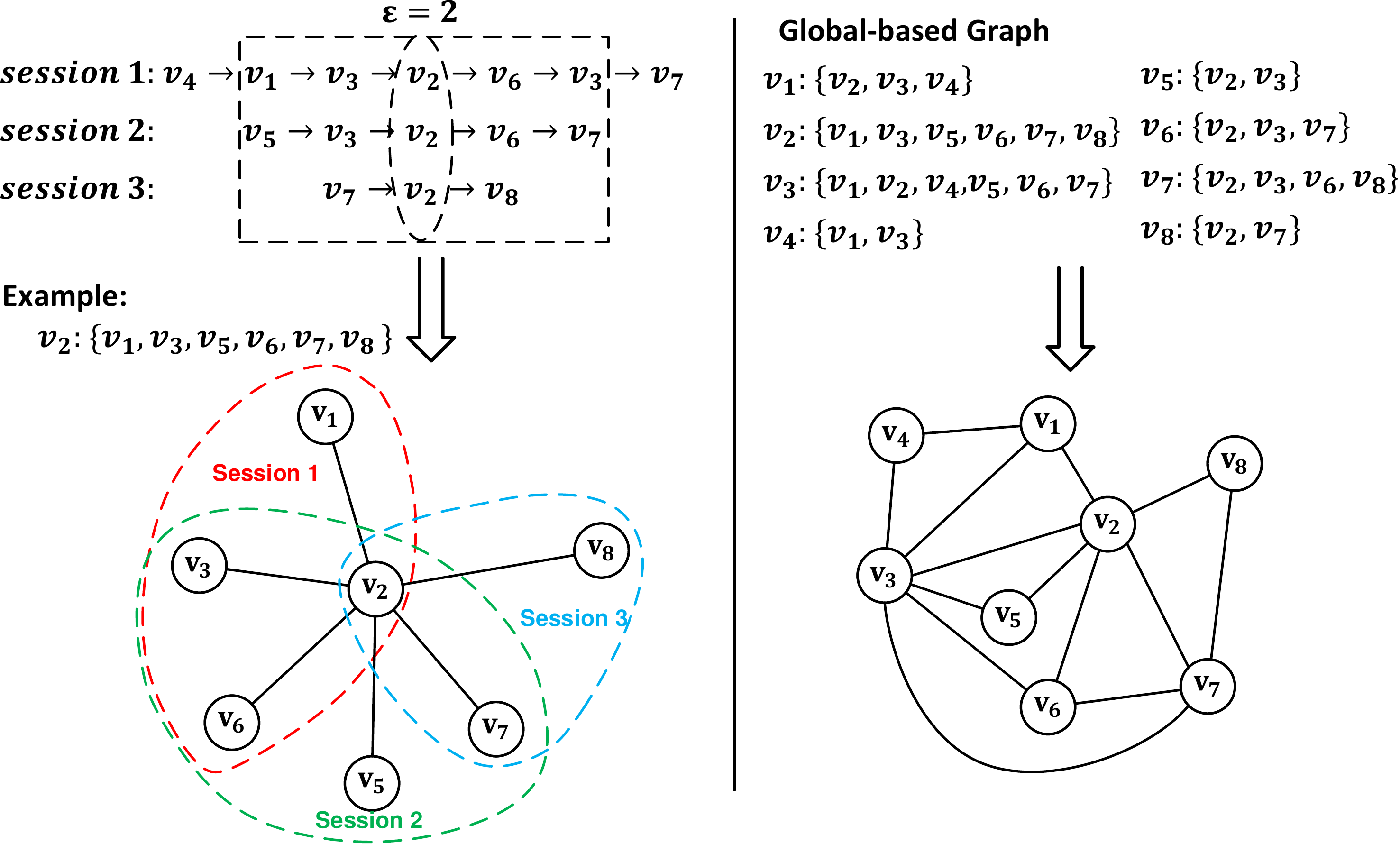}
  \end{minipage}
  \label{globalGraph}
 }
 \caption{Illustrations of construction of \emph{session} graph and  \emph{global} graph.}
\end{figure}

\subsubsection{Global Graph Model}
Similar to conventional recurrent neural network (\eg RNN \cite{li2017neural}) based approaches,
%
\emph{session} graph can efficiently capture
sequential \emph{intra}-relations
to learn \emph{session}-level item embeddings.
%
%
%
However, it neglects the complicated \emph{inter}-relations of items over sessions,
\eg
the item-transition information from other sessions, which is called \textbf{\emph{global}-level} item transition information.


\paratitle{Global-level Item Transition Modeling}.
Here,
we  take into account \emph{global}-level item transitions
for \emph{global}-level item representation learning, via integrating all pairwise item transitions over sessions.
%
As such, we propose a novel \emph{global} graph model for learning \emph{global}-level item embeddings,
which breaks down sequence independence assumption with linking all pairs of items
based on pairwise transitions over all sessions (including the current one).
Next, we firstly introduce a new definition (\ie~{$\varepsilon$-neighbor set}) for modeling global-level item transition,
and then give the definition of \emph{global} graph.

\begin{definition}
\label{definition-1}
\textbf{$\varepsilon$-Neighbor Set}~($\mathcal{N}_{\varepsilon}({v})$).
\emph{For any item $v^{p}_i$ in session $S_p$, the $\varepsilon$-neighbor set of $v^{p}_i$ indicates a set of items,
each element of which is defined as follows,
{
\begin{equation}
\begin{split}
    \mathcal{N}_{\varepsilon}(v^{p}_i)= 
    \{ 
    v^{q}_{j} | 
    & v^{p}_{i}=v^{q}_{i^{'}}\in S_p\cap S_q; v^{q}_{j}\in S_q; \\
    & j \in [i^{'}-\varepsilon,i^{'}+\varepsilon]; S_p\neq S_q
    \},
\end{split}    
\end{equation}
}
where $i^{'}$ is the order of item $v^{p}_i$ in session $S_q$,
$\varepsilon$ is a hyperparameter to control the scope of modeling of item-transition  between $v^{p}_i$
and the items in $S_q$.
Note that, parameter $\varepsilon$ favors the modeling of short-range item transitions
over sessions, since it is helpless (even noise, \eg irrelevant dependence) for
capturing the \emph{global}-level item transition information if beyond the scope ($\varepsilon$).
}
\end{definition}
According to Definition~\ref{definition-1}, for each item $v_i\in V$, \textbf{\emph{global}-level item transition}
is defined as
\begin{math}
  \{(v_i,v_j)|v_i,v_j\in V; v_j\in \mathcal{N}_{\varepsilon}({v_i})\}.
\end{math}


\paratitle{Global Graph}.
Global graph aims to capture the \emph{global}-level item transition information, which will be used
to learn item embeddings over all sessions.
%
%
Specifically, the  \emph{global} graph is built based on $\varepsilon$-neighbor sets
of items in all sessions.
Without loss of generality,
\emph{global} graph is defined as follows,
%
%
let $\mathcal{G}_{g} = ( \mathcal{V}_{g}, \mathcal{E}_{g} )$ be the \textbf{\emph{global}} graph,
where $\mathcal{V}_{g}$ denotes the graph node set that contains all items in $V$,
and
\begin{math}
\mathcal{E}_{g}=\{e^{g}_{ij}|(v_i,v_j);v_i\!\in\! V, v_j\!\in\! \mathcal{N}_{\varepsilon}({v_i})\}
\end{math}
indicates the set of edges, each corresponding to two pairwise items from all the sessions.
Figure~\ref{globalGraph} shows an example of building the \emph{global} graph with $\varepsilon=2$.
To distinguish the importance of $v_i$'s neighbors ($\mathcal{N}_{\varepsilon}({v_i})$),
we treat the co-occurrence over sessions of node $v_i$ and its neighbor node $v_j\in \mathcal{N}_{\varepsilon}({v_i})$
as the weights of the corresponding edge.
%

\paratitle{Remark}.
(i) The definition of the neighbors\footnote{We do not distinguish $\mathcal{N}_{\varepsilon}({v})$ and $\mathcal{N}^{g}_{v}$ when the context is clear and discriminative.} (\ie $\mathcal{N}^{g}_{v}$) of item $v$ on graph $\mathcal{G}_{g}$ is same as $\mathcal{N}_{\varepsilon}(v)$;
(ii) $\mathcal{G}_{g}$ is an \emph{undirected} \emph{weighted} graph as $\varepsilon$-neighbor set is undirected;
and (iii)  Each item in $V$ is encoded into an unified embedding space at time-step $t$, \ie $\vec{h}^{t}_{i}\in\mathbb{R}^{d}$~($d$ indicates the dimension of item embedding), which is feed with an initialization embedding $\vec{h}^{0}_{i}\in \mathbb{R}^{|V|}$, here we use \emph{one-hot} based embedding and it is transformed into $d$-dimensional latent vector space by using a trainable matrix $\vec{W}_{0}\in \mathbb{R}^{d \times |V|}$.

\section{The Proposed Method}

\begin{figure*}[t]
    \centering
    \subfloat[Basic GNN-based Framework.]{
    \begin{minipage}[b]{\linewidth}
    \centering
    \includegraphics[width=0.8\linewidth]{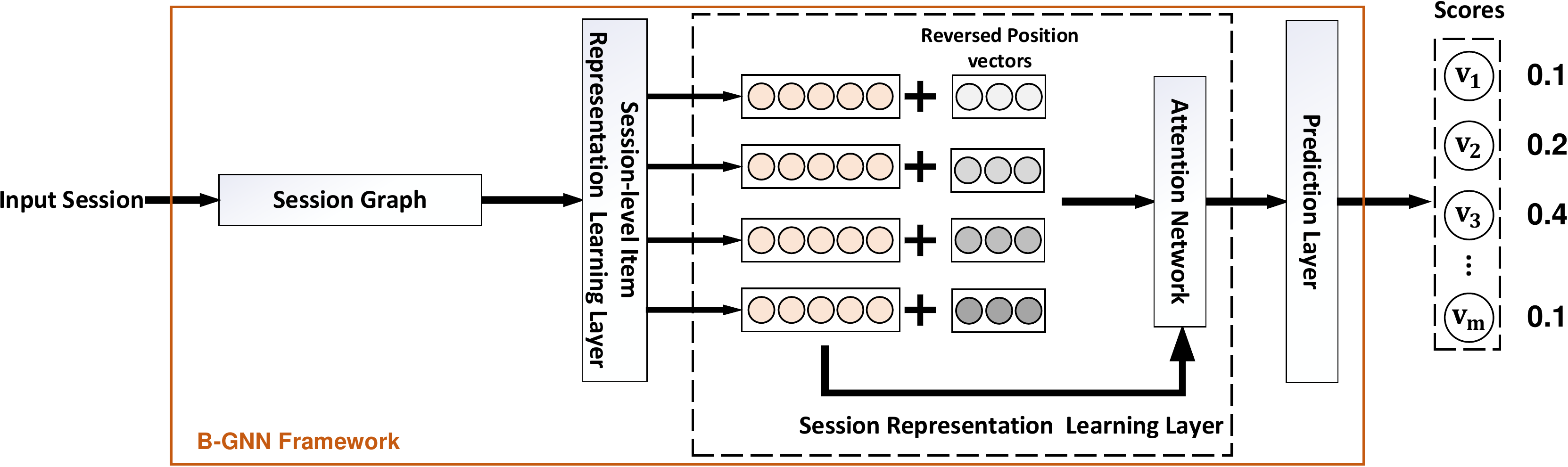}
    \end{minipage}
    \label{fig:basicModule}
    }

    \subfloat[Fusion-based Model.]{
    \centering
    \includegraphics[width=0.5\linewidth]{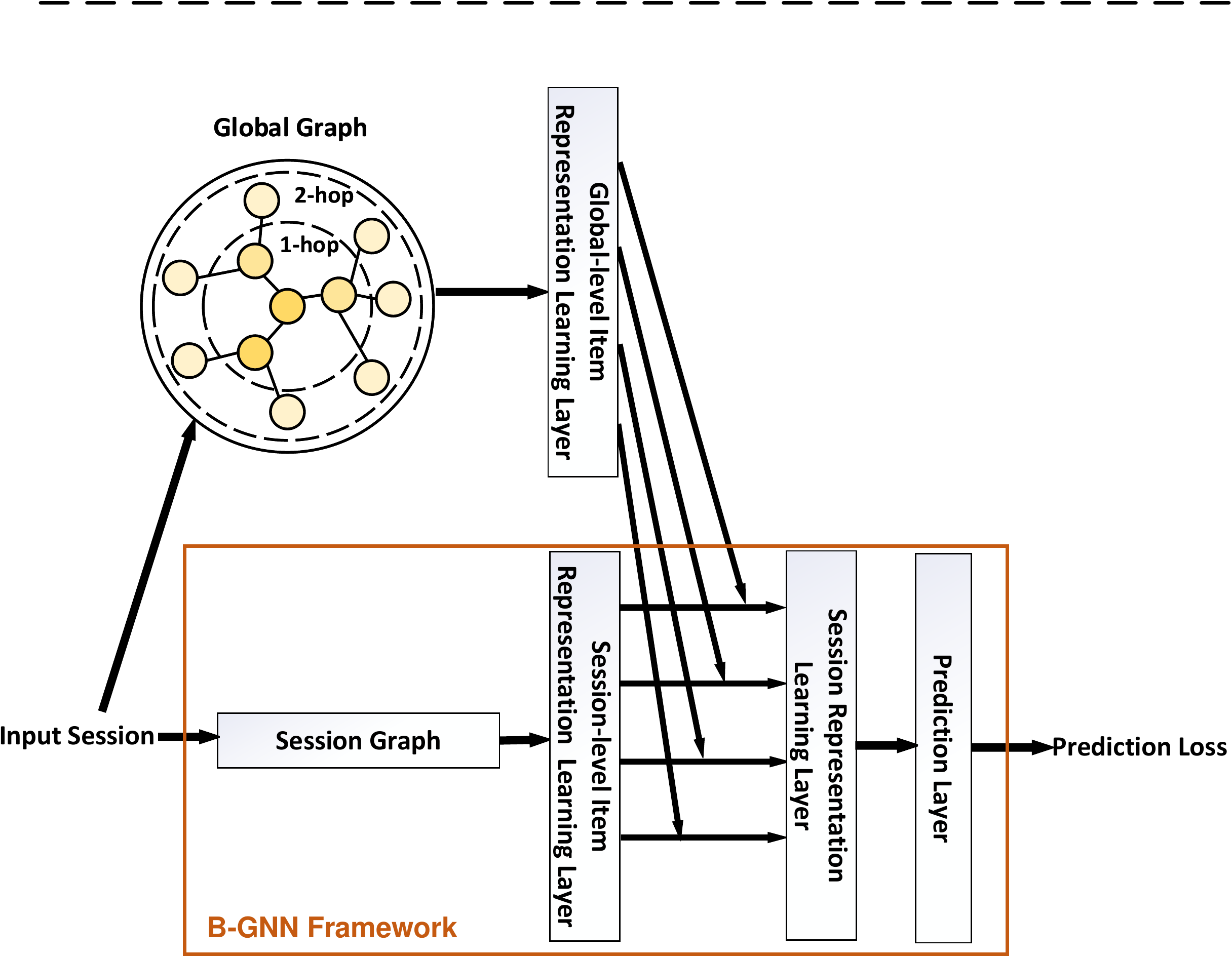}
    \label{fig:AggVersion}
    }
    \subfloat[Constrained-based Model.]{
    \centering
    \includegraphics[width=0.5\linewidth]{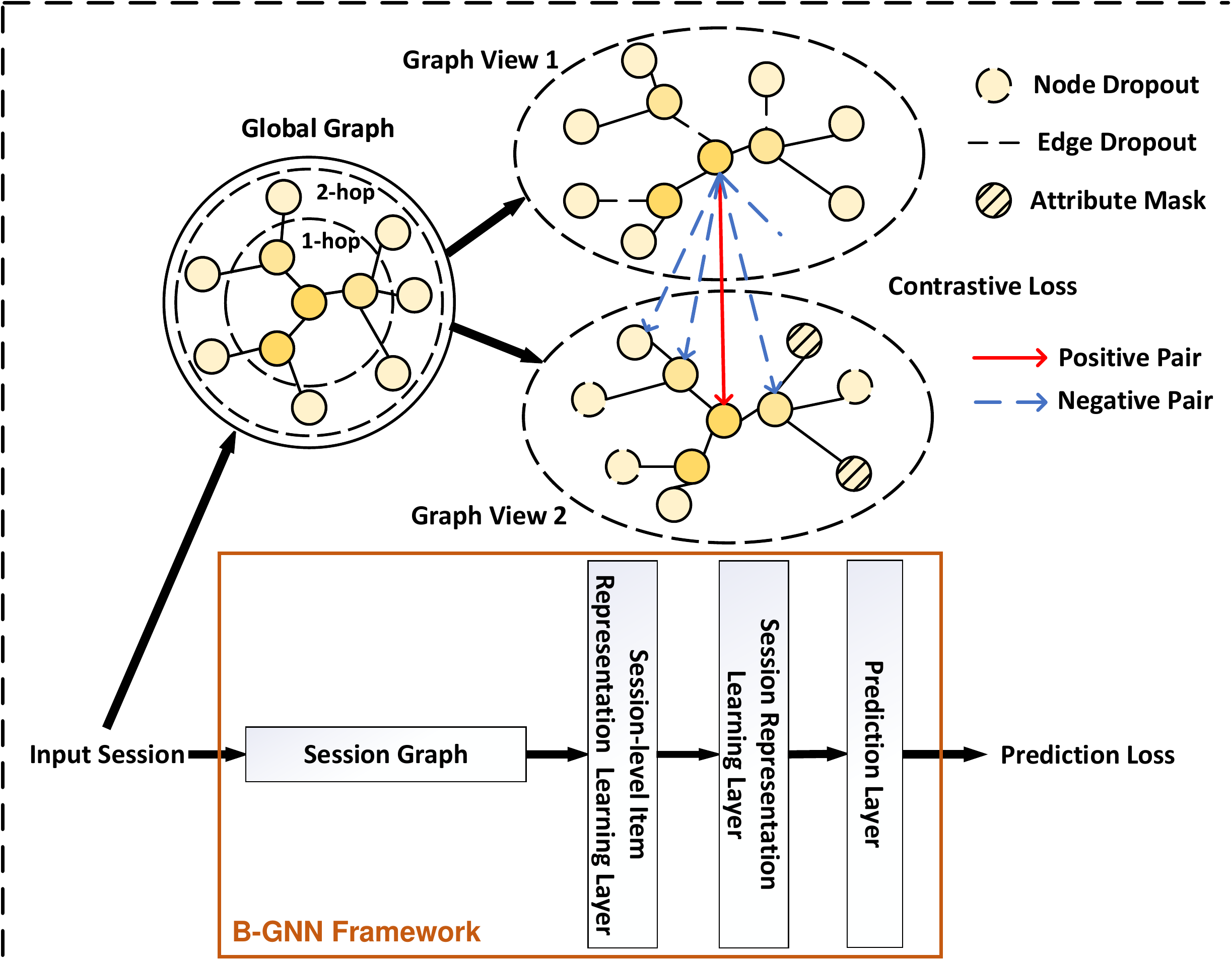}
    \label{fig:EmbVersion}
    }
    \vspace{-8pt}\caption{An overview of the proposed framework. (a) \textbf{Basic GNN-based framework}, which utilizes graph neural networks to exploit the item transitions in session graph. (b) \textbf{Fusion-based model}, where the neighbors of item in global graph are incorporated into current session by a session-level attention mechanism. (c) \textbf{Constrained-based model}, which introduces the global transition information by preserving the global proximity in the embedding space.}
  \label{fig:frameWork}
\end{figure*}

To leverage the different levels (\eg \emph{local}-level and \emph{global}-level) information
for SBR,
we first propose a basic GNN-based framework (\ie \textsf{B-GNN}, as shown in Fig.~\ref{fig:basicModule})
in Section \ref{sec-BGNN}.
Based on the \textsf{B-GNN} framework,
we also propose a novel approach (named \textbf{S}ession-based \textbf{R}ecommendation with \textbf{G}lobal \textbf{I}nformation, \textsf{SRGI})
with two different variants in Section \ref{sec-srgi-fm} and Section \ref{sec-srgi-cm},
which is capable of leveraging the \emph{global} item-transition information
from two different aspects:
(a) \textsf{SRGI-FM}~(\rf Fig.~\ref{fig:AggVersion}), it incorporates the global item-transitions information into the learning process of session-level item representation;
and 
(b) \textsf{SRGI-CM}~(\rf Fig.~\ref{fig:EmbVersion}), it treats the global-level item-transition information
as a constraint to ensure the learnt item embeddings are consistent with the 
structure of the global graph.
Next, we will illustrate each part in detail.

\subsection{Basic GNN-based Framework~(B-GNN)}
\label{sec-BGNN}
In this section, we first propose a basic GNN-based model (called \textsf{B-GNN}, as shown in Fig. \ref{fig:basicModule}),
which solely utilizes three different types (\ie \emph{in}, \emph{out} and \emph{in-out}, as shown in Fig. \ref{sessionGraph}) of session-level item transition information for session-based recommendation.
Specifically, \textsf{B-GNN} consists of three sub-components, \ie session-level item representation learning layer, session representation learning layer and prediction, and we will be detailed in the following sub-sections, respectively.

\subsubsection{Session-level Item Representation Learning layer}
To learn the session-level item representation,
we focus on how to enrich the representation of each item
with the help of its $1$-hop neighbors, via exploring the pairwise item-transitions within the current session.


\paratitle{Information Propagation}:
For each node $v_i$ of session $S$,
we consider three different types of relations between $v_i$ and its neighbors $\mathcal{N}^{s}_{v_i}$,
\ie in-coming neighbor, out-coming neighbor and in-out coming neighbor
which are denoted by $\mathcal{N}^{s, in}_{v_i}$, $\mathcal{N}^{s, out}_{v_i}$
and $\mathcal{N}^{s, in-out}_{v_i}$
respectively.
To calculate the importance of different neighbors,
we treat such three kinds of neighbors differently during the propagation of information,
which are first calculated based on mean pooling,
namely,
\begin{equation}
\begin{split}
    \vec{h}_{N_{v_i}^{s, in}} &= \text{MeanPooling}({\vec{h}_{v_j} | v_j \in \mathcal{N}_{v_i}^{s, in}}) \\
    \vec{h}_{N_{v_i}^{s, out}} &= \text{MeanPooling}({\vec{h}_{v_j} | v_j \in \mathcal{N}_{v_i}^{s, out}})\\
    \vec{h}_{N_{v_i}^{s, in-out}} &= \text{MeanPooling}({\vec{h}_{v_j} | v_j \in \mathcal{N}_{v_i}^{s, in-out}})
\end{split}
\end{equation}

Then, the neighbor representation ($\vec{h}_{N_{v_i}^{s}}$) of node $v_i$ is obtained based on the concatenation
of $\vec{h}_{N_{v_i}^{s, in}}$, $\vec{h}_{N_{v_i}^{s, out}}$ and $\vec{h}_{N_{v_i}^{s, in-out}}$,
namely,
\begin{equation}\label{concatenation-bi}
  \vec{h}_{N_{v_i}^{s}} = [\vec{h}_{N_{v_i}^{s, in}} || \vec{h}_{N_{v_i}^{s, out}} || \vec{h}_{N_{v_i}^{s, in-out}}],
\end{equation}
where $||$ denotes concatenation operation.
In particular,
different from SR-GNN~\cite{wu2019session} and FGNN~\cite{qiu2019rethinking}, 
here we consider three kinds of relations on session graph and use mean pooling in our propagation layer to reduces the training parameters for avoiding over-fitting.


\paratitle{Information Aggregation}:
The \textbf{\emph{session-level}} item representation
is generated by aggregating
the embeddings of item $v_i$
and its neighbors $\vec{h}_{N_{v_i}^{s}}$,
which is computed via a fully connected layer,
\begin{equation}
    \vec{h}^{s}_{v_i} = \text{tanh} (\matrix{W}_1 \vec{h}_{v_i} + \matrix{W}_2 \vec{h}_{N_{v_i}^{s}} + \vec{b}_1 ),
\end{equation}
where $\matrix{W}_1 \in \mathbb{R}^{d \times d}, \matrix{W}_2 \in \mathbb{R}^{d \times 3d}$ and $\vec{b}_1 \in \mathbb{R}^{d}$ are trainable parameters. The new representations $\vec{h}^{s}_{v_i}$ of each item is aggregated by the features of item itself and its neighbors in the current session.

\subsubsection{Session Representation Learning Layer}
Based on the learnt item representations, we now present how to obtain the session representations.
Note that the contribution of different items to the next prediction is not equal.
Intuitively, the items clicked later in the session are more representative of the user's current preferences. Moreover, it is important to find the main purpose of the user and filter noise in current session \cite{li2017neural}. Hence we incorporate the reversed position information and session information to make a better prediction.
After feeding a session sequence into graph neural networks, we can obtain the representation of the items involved in the session, \ie $\matrix{H} = \left[ \vec{h}_{v^s_1}^{\prime}, \vec{h}_{v^s_2}^{\prime}, ..., \vec{h}_{v^s_l}^{\prime} \right]$\footnote{Here $\vec{h}_{v^s_1}^{\prime} = \vec{h}^{s}_{v_i}$ for B-GNN.}. We also use a learnable position embedding matrix $\matrix{P} = \left[ \vec{p}_{1}, \vec{p}_{2}, ..., \vec{p}_{l} \right]$, where $\vec{p}_{i} \in \mathbb{R}^d$ is a position vector for specific position $i$ and $l$ is the length of the current session sequence. The position information is integrated through concatenation and non-linear transformation:
\begin{equation}
\label{eq-reverse-position}
\vec{z}_{i} = \text{tanh} \left( \matrix{W}_{3}\left[ \vec{h}_{v^s_i}^{\prime} \parallel \vec{p}_{l-i+1} \right] + \vec{b}_{2} \right),
\end{equation}
where parameters $\matrix{W}_{3} \in \mathbb{R}^{d \times 2d}$ and $\vec{b}_{2} \in \mathbb{R}^{d}$ are trainable parameters.
Here reversed position $l-i+1$ can be regard as the distance between the current item and the predicted item, which contains more effective information than forward position $i$.
Next, the corresponding weight is learned through a soft-attention mechanism:
\begin{equation}
\beta_{i} = \vec{q}_2^{\top} \sigma \left( \matrix{W}_{4} \vec{z}_{i} + \matrix{W}_{5} \vec{s}^{\prime} + \vec{b}_{3} \right),
\end{equation}
where $\matrix{W}_{4}, \matrix{W}_{5} \in \mathbb{R}^{d \times d}$ and $\vec{q}_2, \vec{b}_{3} \in \mathbb{R}^{d}$ are learnable parameters and $\vec{s}^{\prime}$ denotes session information, which is obtained by sum pooling (\ie $\vec{s}^{\prime} = \sum_{i=1}^l \vec{h}_{v^s_i}^{\prime}$). Finally, the session representation can be obtained by linearly combining the item representations:
\begin{equation}
\vec{S} = \sum_{i=1}^l \beta_{i} \vec{h}_{v^s_i}^{\prime}.
\end{equation}

The session representation $\vec{S}$ is constructed by all the items involved in the current session, where the contribution of each item is determined not only by the information in the session graph, but also by the chronological order in the sequence.

\subsubsection{Objective Function}
Based on the obtained session representations $\vec{S}$, the final recommendation probability for each candidate item based on their initial embeddings as well as current session representation.
Here, $\ell_2$-norm is employed
to normalize  session representation and item representations (\ie $\hat{\vec{S}} = \frac{\vec{S}}{||\vec{S}||_2}$ and $ \hat{\vec{h}}_{v}  = \frac{\vec{h}_v}{||\vec{h}_v||_2}$)
for avoiding popular bias \cite{gupta2019niser}.
Then,
we use inner-product and apply softmax function to obtain the output $\hat{y}$:
\begin{equation}
\hat{y}_i = \text{Softmax} \left( \alpha \hat{\vec{S}}^{\top} \hat{\vec{h}}_{v_i} \right),
\end{equation}
where $\alpha$ is a scale coefficient for better convergence \cite{gupta2019niser, wang2017normface} and $\hat{y}_{i} \in \hat{y}$ denotes the probability of item $v_{i}$ appearing as the next-click in the current session.

The loss function is defined as the cross-entropy of the prediction results $\hat{y}$:
\begin{equation}
\label{eq-obj}
\mathcal{L}_S = -\sum_{i=1}^{m} \vec{y}_{i} \log \left(\hat{y}_{i}\right)+\left(1-\vec{y}_{i}\right) \log \left(1-\hat{y}_{i}\right),
\end{equation}
where $\vec{y}$ denotes the one-hot encoding vector of the ground truth item.

\subsection{Fusion-based Model}
\label{sec-srgi-fm}


In this section, we propose a novel fusion-based model under the \textsf{B-GNN} framework,
\ie a variant of our proposed model \textsf{SRGI} (\eg \textsf{SRGI-FM}, as shown in Fig. \ref{fig:AggVersion}).
We will detail how to incorporate the global neighbor features into current session,
which is built based on the architecture of graph convolution network \cite{kipf2016semi}.
Here we first describe a single layer consisting of two components: \emph{fusion-based information propagation} and \emph{information aggregation}, and then show how to generalize it to multiple layers.

\paratitle{Fusion-based Information Propagation}:
To obtain the first-order (\ie 1-hop neighbors in the global graph) neighbor's features of item $ v $,
one straightforward solution is to use mean pooling method \cite{hamilton2017inductive}.
However, not all of items in $ v $'s $\varepsilon$-neighbor set are relevant to the user preference of the current session,
and thus we consider to utilize a session-aware attention to distinguish the importance of items in ($\mathcal{N}_{\varepsilon}(v)$).
Therefore, each item in $\mathcal{N}_{\varepsilon}(v)$ is linearly combined according to the session-aware attention score,
\begin{equation}
\vec{h}_{\mathcal{N}^g_{v_i}} = \sum_{v_j \in \mathcal{N}^g_{v_i}} \pi (v_i, v_j) \vec{h}_{v_j} ,
\end{equation}
where $\pi (v_i, v_j)$ estimates the importance weight of different neighbors. Intuitively, the more consistent an item is to the preference of the current session, the more important this item is to the recommendation. Therefore, we implement $\pi (v_i, v_j)$
based on the principle of attention network \cite{velivckovic2017graph}:
\begin{equation}
\pi (v_i, v_j) = \vec{q}^T_1 \text{LeakyRelu} \big ( \matrix{W}_6 [ (\vec{s}^{\prime} \odot \vec{h}_{v_j}) \| w_{ij} ] \big ).
\end{equation}
where $\|$ indicates concatenation operation; $w_{ij} \in \mathbb{R}^{1}$ is the weight of edge $(v_i, v_j)$ in global graph; $\matrix{W}_6 \in \mathbb{R}^{d+1 \times d+1}$ and $\vec{q}_1 \in \mathbb{R}^{d+1}$ are trainable parameters; $\odot$ indicates element-wise product.

Different from mean pooling,
in our model the propagation of information depends
on the affinity between $\vec{s}^{\prime}$ and $ v_{j} $, which means neighbors that match the preference of current session will be more favourable,
and we then normalize the coefficients across all neighbors connected with $v_i$ by adopting the softmax function:
\begin{equation}
\label{eq-pai}
\pi (v_i, v_j) = \frac{\exp \big( \pi (v_i, v_j) \big)}{\sum_{v_k \in \mathcal{N}^g_{v_i}} \exp \big( \pi (v_i, v_k) \big)}.
\end{equation}

As such, Eq.~(\ref{eq-pai}) is capable of
assigning high attention scores to the important global-level 1-hop neighbors
of items in the current session.

\paratitle{Fusion-based Information Aggregation}: The final step is to aggregate the item representation $\vec{h}_{v}$ and its neighborhood representation $ h^g_{\mathcal{N}_v} $, we implement aggregator function \text{agg} as follows,
\begin{equation}
\vec{h}^{g}_{v} = \text{Relu} \big( \matrix{W}_{7} [ \vec{h}_{v} || \vec{h}_{\mathcal{N}^g_v} ]  \big),
\end{equation}
where $\matrix{W}_{7} \in \mathbb{R}^{d \times 2d}$ is transformation weight.

Through a single aggregator layer, the representation of an item is dependent on itself and its immediate neighbors. We could explore the high-order connectivity information through extending aggregator from one layer to multiple layers, which allows more global information to be incorporated into the current representation. We formulate the representation of an item in the $k$-th steps as:
\begin{equation}
\vec{h}^{g, (k)}_{v} = \text{agg} \big ( \vec{h}^{(k-1)}_{v}, \vec{h}^{(k-1)}_{\mathcal{N}^g_{v}} \big ),
\end{equation}
$\vec{h}^{(k-1)}_{v}$ is representation of item $v$ which is generated from previous information propagation steps, $\vec{h}^{(0)}_{v}$ is set as $\vec{h}_{v}$ at the initial  propagation iteration. In this way, the $k$-order representation of an item is a mixture of its initial representations and its neighbors up to $k$ hops away. This enables more effective messages to be incorporated into the representation of the current session.

After obtaining the representation of the $\vec{h}^{g, (k)}_{v}$, the new representation of each item can be obtained as follow,
\begin{equation}
      \vec{h}^{\prime}_v = \vec{h}^{g, (k)}_{v} + \vec{h}^{s}_{v}.
\end{equation}

Consequently, we obtain the new representation for each item of the current session,
which integrate both session-level
and global-level item-transition information
and thus the recommendation prediction can be implemented based on Eq. (\ref{eq-reverse-position})-(\ref{eq-obj})
via using the new item embedding $\vec{h}^{\prime}_v$.


\subsection{Constrained-based Model}
\label{sec-srgi-cm}
Recall that the strategy of building the global graph
is based on the co-occurrence of pairwise items over sessions,
namely, pairwise items with high frequency co-occurrences are remained while filtering the ones with low-frequency co-occurrences.
To this end, in this section we propose another variant of our proposed model \textsf{SRGI} (\ie \textsf{SRGI-CM}, as shown in Fig. \ref{fig:EmbVersion}),
which is different from \textsf{SRGI-FM} that aims to integrate different levels of item-transitions for learning item representation,
the principle of \textsf{SRGI-CM} is to ensure the learnt item embeddings are consistent with global-level item-transition information, which is called \emph{global proximity}
that can be modeled by graph contrastive learning from two different aspects:
(i) Preserving the local structural patterns. An item and its global neighbors often have high relevancy (\eg milk and bread) due to high frequency item-transitions over sessions,
and thus should be assigned to the proximal vectors in the learnt embedding space;
and, ii) Enforcing the discrimination for different items. The representation of different nodes should be discriminating in the learned embedding space.
Next, we show the process of performing contrastive learning on the global graph.

\paratitle{Graph Data Augmentation.} 
Given the global graph $\mathcal{G}_{g} = ( \mathcal{V}_{g}, \mathcal{E}_{g} )$, we first construct two correlated graph views based on data augmentation. Graph augmentation aims to create novel and realistically rational data via certain transformations on the original graph. Different graph augmentations for graph provide different contexts for each node, which is the key component of graph contrastive learning. Following previous graph contrastive learning approaches \cite{you2020graph,zhu2020graph}, three general data augmentations are used for creating novel and realistically global graphs:

\begin{enumerate} 
    \item Node dropping. It will randomly discard a certain portion of nodes in the global graph and also remove their corresponding connections, where the dropping probability for each node follows a uniform distribution.
    \item Edge dropping. Similar to node dropping, edge dropping will randomly remove certain portion of edges in the original graph.
    \item Attribute masking. It masks a certain fraction of dimensions of node features with zeros.
\end{enumerate}

\paratitle{Contrastive Learning.}
After employing data augmentations we can obtain two correlated graph views which are both corrupted from the original global graph. We denotes two graph views as $\mathcal{G}^1_{g} = ( \mathcal{V}^1_{g}, \mathcal{E}^1_{g} )$ and $\mathcal{G}^2_{g} = ( \mathcal{V}^2_{g}, \mathcal{E}^2_{g} )$, where $\mathcal{V}_{*}$ and $\mathcal{E}_{*}$ are the node sets and the edge sets for two graph views. Then a GNN-based encoder $f(\cdot, \cdot)$ is applied here to extract the features for each node in the graph:
\begin{equation}
\begin{split}
    \vec{o}^1 &= f ( \mathcal{V}^1_{g}, \mathcal{E}^1_{g} ) \\
    \vec{o}^2 &= f ( \mathcal{V}^2_{g}, \mathcal{E}^2_{g} )
\end{split}
\end{equation}
where $\vec{o}^*_i \in \matrix{o}^*$ is the learnt node representation for node $i$ in graph view $\mathcal{G}^*_{g}$. Contrastive learning does not apply any constraint on GNN architecture \cite{you2020graph}, and here we choose GCN~\cite{kipf2016semi} for its low computational cost and stability.
After obtaining the node representations $\vec{o}^1$ and $\vec{o}^2$, we use a projection function (\ie a single fully connected layer) to transform the learnt node representation to another latent space where the contrastive loss is calculated,
\begin{equation}
    \vec{z} = \text{Relu} (\matrix{W}_8 \vec{o} + \vec{b}_4).
\end{equation}

Finally, the contrastive loss is employed to maximize the node-level agreement while enforcing the representations of different nodes discriminate in the latent space. For the node $i$ in the graph view $\mathcal{G}^1_{g}$, only the corresponding node in the graph view $\mathcal{G}^2_{g}$ is regarded as positive samples, while all other nodes are treated as negative samples. The contrastive loss for the global graph is shown as follow:
\begin{equation}
\label{eq-lp}
\mathcal{L}_C = \sum\limits^{N}_{i=1}log \frac{e^{\theta(\vec{z}^1_{i}, \vec{z}^2_{i})}}{e^{\theta(\vec{z}^1_{i}, \vec{z}^2_{i})} + \sum\limits^{N}_{k=1}\vec{I}_{[k \neq i]} e^{\theta(\vec{z}^1_{i}, \vec{z}^2_{k})}
+ \sum\limits^{N}_{k=1}\vec{I}_{[k \neq i]} e^{\theta(\vec{z}^1_{i}, \vec{z}^1_{k})}
}
\end{equation}
where $\vec{I}_{[*]}$ is an indicator function and $\theta(a, b) = cos(a, b)$ denotes the cosine similarity function. The contrastive loss $\mathcal{L}_C$ is aim to: i) Maximizing the consistency between the corresponding node in different graph views, which will force each node's features to be proximal with its neigbhors' features and thus the learned node representations will be more robust; and ii) Minimizing the agreement between different nodes, which lead to the discrimination of different nodes' representations in the latent embedding space.

To learn the item transitions during the session and preserve the global proximity, the final loss function of SRGI-CM combines prediction loss and contrastive loss,
\begin{equation}
    \mathcal{L}_{G} = \mathcal{L}_S + \lambda_C \mathcal{L}_C,
\end{equation}
where $\lambda_C$ is a trade-off parameter to control the impact of global proximity.

\subsection{Discussion}
As we show the detail of SRGI-FM and SRGI-CM, the common features of these two models is that they are both designed based on the technology of GNN and the proposed global graph. The global item transition is important for the modeling of the item-transition within the current session. By exploring the node features and structural information in the global graph, the global context information is introduced to enrich the representation of each item. And the experiment results demonstrate that both SRGI-FM and SRGI-CM show promising improvements compared with B-GNN.

And the main difference between SRGI-FM and SRGI-CM is reflected in two aspects:
(i) The fusion-based model SRGI-FM and constrained-based model SRGI-CM present two different ways to incorporate the effective global information into the current session. SRGI-FM uses attention-based GNN to directly incorporate the neighbor features from the global graph into the current session, while SRGI-CM focuses on the global graph structure and leverages  graph contrastive learning to constrain the representation of each item in the latent space.
(ii) Comparing with SRGI-FM, SRGI-CM needs to employ data augmentation to create different graph views based on the global graphs when in the training process. However, in the inferring process, SRGI-CM does not need the global graph as input or compute the contrastive loss, while SRGI-FM still needs to incorporate the features from the global graph into the current session. This means SRGI-CM requires more time to process the data during the training process, while in the process of inference, the computation cost of SRGI-CM will be much smaller than SRGI-FM.

\section{Experiments}

We have conducted extensive experiments to evaluate the accuracy of the proposed method by answering the following four key research questions:

\begin{itemize} [itemindent = 15pt]
\setlength{\itemsep}{3pt}

\item \noindent \textbf{RQ1}: Does two versions of SRGI outperform state-of-the-art SBR baselines in real world datasets?

\item \noindent \textbf{RQ2}: Does global graph information improve the performance of SRGI? What is the impact of different hyper-parameter on learning global item transitions?

\item \noindent \textbf{RQ3}: How does the basic GNN framework~(\ie B-GNN) perform compared with other session-level item representation learning methods?

\item \noindent \textbf{RQ4}: Does reversed position aware session encoder show effective for session-based recommendation?

\end{itemize}

\subsection{Datesets and Preprocessing}
To fully evaluate the effectiveness of our proposed method, we employ three real-world datasets in the experiment.

$Diginetica$\footnote{https://competitions.codalab.org/competitions/11161} is from CIKM Cup 2016, which consisting of typical transaction data for five month from e-commerce website. Following \cite{wu2019session}, we set the sessions of last week (latest data) as the test data, and the remaining historical data for training.

$Tmall$\footnote{https://tianchi.aliyun.com/dataset/dataDetail?dataId=42} comes from IJCAI-15 competition, which contains anonymized user's shopping logs on Tmall online shopping platform. Since the amount of items of Tmall is extremely large, we use the first 120,000 of the sessions for train and test. The sessions of last 100 seconds are set as the test data, and the remaining historical data for training.

$Nowplaying$\footnote{https://dbis-nowplaying.uibk.ac.at/\#nowplaying} comes from music-related tweets~\cite{ismm14}, which describes the music listening behavior of users. We set the sessions of last two months as the test data, and the remaining historical data for training.

Following \cite{wu2019session,ijcai2019-547}, we conduct preprocessing step over the three datasets. More specifically, sessions of length 1 and items appearing less than 5 times were filtered across all the three datasets. For Tmall, sessions longer than 40 were also filtered~\cite{wang2019collaborative}. Furthermore, for a session $ S = \left[ s_{1}, s_{2}, ..., s_{n} \right]$, we generate sequences and corresponding labels by a sequence splitting preprocessing, i.e., ($\left[ s_1 \right], s_2$), ($\left[ s_1, s_2 \right], s_3$), ..., ($\left[ s_1, s_2,..., s_{n-1} \right], s_n$) for both training and testing across all the three datasets. The statistics of datasets, after preprocessing, are summarized in Table \ref{tab:statistic}.
{
\begin{table}[t]
	\centering
	\caption{Statistics of the used datasets.}
	\label{tab:statistic}
	\begin{tabular}{l|r|r|r}
	\hline
		{ \text{Dataset} } & \text{Diginetica} & \text{Tmall} & \text{Nowplaying}\\
		\hline
		{ \text{\# train sessions} }  & 719,470 & 351,268 & 825,304 \\
		\hline
		{ \text{\# test sessions} }   & 60,858 & 25,898 & 89,824 \\
		\hline
		{ \text{\# items} }  & 43,097 & 40,728 & 60,417 \\
		\hline
		{ \text{avg. len.} } & 5.12 & 6.69 & 7.42 \\
	\hline
\end{tabular}
\end{table}
}

\subsection{Evaluation Metrics}
We adopt two widely used ranking based metrics: \textbf{P@N} and \textbf{MRR@N} by following previous work\cite{liu2018stamp, wu2019session}.

\textbf{P@N}~(Precision)\cite{wu2019session}: The P@N score is typically used as a measure of accuracy. It represents the proportion of correctly recommended items in 
top $N$ recommended item list, which is defined as:
\begin{equation}
P@N = \frac{n_{hit}}{n_{test}},
\end{equation}
where $n_{test}$ denotes the number of test data and $n_{hit}$ denotes the number of  the target items appearing in the  of top $N$ recommended items.

\textbf{MRR@N}~(Mean Reciprocal Rank)\cite{liu2018stamp}: The MRR@N score is the average of reciprocal rank of the correctly-recommended items. The reciprocal rank is set to zero if the rank exceeds $N$,
\begin{equation}
MRR@N = \frac{1}{n_{test}} \sum \frac{1}{\text{Rank}(v_{target})}.
\end{equation}
MRR is a normalized score in the range of $[0, 1]$, and a larger \text{MRR} value means that correct recommendations are in the top of the ranking list.

Here, we choose $N = 20$ for both P@N and MRR@N, as recommendation systems should focus on top ranked items.

\subsection{Baseline Algorithms}
We compare our method with classic methods as well as state-of-the-art models. The following baseline models are evaluated.

\paratitle{POP}: It recommends top-$N$ frequent items of the training set.

\paratitle{Item-KNN}\cite{sarwar2001item}: It recommends items based on the similarity between items of the current session and items of other ones.

\paratitle{FPMC}\cite{rendle2010factorizing}: It combines the matrix factorization and the first-order Markov Chain for capturing both sequential effects and user preferences. By following the previous work, we also ignore the user latent representations when computing recommendation scores.

\paratitle{GRU4Rec}\footnote{https://github.com/hidasib/GRU4Rec} \cite{hidasi2015session}:
It is a RNN-based model that uses Gated Recurrent Unit (GRU) to model user sequences.

\paratitle{NARM}\footnote{https://github.com/lijingsdu/sessionRec\_NARM} \cite{li2017neural}: It improves over \textbf{GRU4Rec} \cite{hidasi2015session} by incorporating attention mechanism into hierarchical RNN for SBR.

\paratitle{STAMP}\footnote{https://github.com/uestcnlp/STAMP} \cite{liu2018stamp}: It employs attention layers to replace all RNN encoders in previous work by fully relying on the self-attention of the last item in the current session to capture the user's short-term interest.

\paratitle{SR-GNN}\footnote{https://github.com/CRIPAC-DIG/SR-GNN} \cite{wu2019session}: It converts sessions into graphs and leverages gated GNN layer to capture the dependencies between items and its context, followed by a self-attention of the last item as STAMP\cite{liu2018stamp} does to obtain the session level representation.

\paratitle{CSRM}\footnote{https://github.com/wmeirui/CSRM\_SIGIR2019} \cite{wang2019collaborative}: It utilizes the memory networks \cite{weston2014memory} to investigate the latest $m$ sessions for better predicting the intent of the current session.


\paratitle{CoSAN}~\cite{ijcai2020-359}: It injects the embedding of neighbor sessions to enrich the item representation and employ multi-head attention to capture the dependencies between items.

\paratitle{GCE-GNN}\footnote{https://github.com/CCIIPLab/GCE-GNN}~\cite{wang2020global}: A state-of-the-art GNN-based model which directly aggregates global information into current session and utilizes attention mechanism to obtain the session representation.

\subsection{Parameter Setup}

Following previous methods \cite{li2017neural}\cite{liu2018stamp}\cite{wu2019session}, the dimension of the latent vectors is fixed  to $100$, and the size for mini-batch is set to $100$ for all models.
We keep the hyper-parameters of each model consistent for a fair comparison.
For CSRM, we set the memory size to 100 which is consistent with the batch size.
For our model, all parameters are initialized using a Gaussian distribution with a mean of $0$ and a standard deviation of $0.1$.
We use the Adam optimizer with the initial learning rate $0.001$, which will decay by $0.1$ after every $3$ epoch. The L2 penalty is set to $10^{-5}$ and the scale coefficient $\alpha$ is set to 12.
Moreover, we set the maximum distance of adjacent items $\varepsilon$ and the number of neighbors to $3$ and $12$, respectively.
In SRGI-FM we use dropout \cite{srivastava2014dropout} to avoid overfitting, the dropout ratio is set to $0.5$ and the graph depth is searched in $\{1, 2\}$.
In SRGI-CM, the graph depth is set to $2$ and the trade-off parameter $\lambda_C$ is searched in $\{10, 50, 100, 150, 100\}$.
All the parameters are searched on a validation set which is a random $10\%$ subset of the training set.

{
\renewcommand
\arraystretch{1.2}
\begin{table*}[ht]
    \centering
    \small
    \caption{The performance of evaluated methods on three datasets.}
    \label{tab:results}
    \begin{tabular}{ccclcclcc}
        \toprule[1pt]
        \multirow{2}{*}{Method} & \multicolumn{2}{c}{Diginetica} & & \multicolumn{2}{c}{Tmall} &  & \multicolumn{2}{c}{Nowpalying} \\
        \cline{2-3} \cline{5-6} \cline{8-9}  \multicolumn{1}{c}{} & P@20    & MRR@20  & & P@20 & MRR@20 & & P@20  & MRR@20 \\
        \hline
        POP       & 1.18  & 0.28  & & 2.00    & 0.90  & & 1.76   & 0.58  \\
        Item-KNN  & 35.75 & 11.57 & & 9.15    & 3.31  & & 14.78  & 4.67  \\
        FPMC      & 22.14 & 6.66  & & 16.06    & 7.32  & & 7.36  & 2.82     \\
        \hline
        GRU4Rec   & 30.79 & 8.22  & & 10.93   & 5.89  & & 7.92   & 4.48  \\
        NARM      & 48.32 & 16.00 & & 23.30   & 10.70  & & 18.59    & 6.93  \\
        STAMP     & 46.62 & 15.13 & & 26.47   & 13.36  & & 17.66    & 6.88  \\
        CSRM      & 48.49 & 17.13 & & 29.46   & 13.96  & & 18.14    & 6.42  \\
        CoSAN     & 51.97 & 17.92 & & 32.68   & 14.09  & & 21.05    & 7.53  \\
        \hline
        SR-GNN    & 50.73 & 17.59 & & 27.57   & 13.72  & & 18.87    & 7.47  \\
        GCE-GNN   & 54.22 & 19.04 & & 33.42   & 15.42  & & 22.37    & 8.40  \\
        \hline
        B-GNN     & 53.93 & 19.05 & & 35.77  & 16.23 & & 22.16 & 8.78 \\
        SRGI-FM & \underline{54.35} & \underline{19.14} & & \underline{36.33}  & \underline{16.42} & & \underline{22.77}   & \textbf{8.87} \\
        SRGI-CM & \textbf{54.71} & \textbf{19.43} & & \textbf{37.28}  & \textbf{17.85} & & \textbf{22.73}   & \underline{8.80} \\
        \bottomrule[1pt]
    \end{tabular}
\end{table*}
}

\subsection{Overall Comparison (RQ1)}

Table \ref{tab:results} reports the experimental results of the state-of-the-art baselines and our proposed model on three real-world datasets, in which the best result of each column is highlighted in boldface. It can be observed that SRGI-FM and SRGI-CM achieve better performance than state-of-the-art baselines across all three datasets in terms of the two metrics, which ascertains the effectiveness of our proposed method.

Among the traditional methods, POP's performance is the worst, as it only recommends the most popular item without learning the preference of the user.
Comparing with POP, FPMC performs better on three datasets, which shows the strength of Markov Chains in modeling sequential data.
Moreover, Item-KNN achieves the best results among the traditional methods on the Diginetica and Nowplaying datasets, which demonstrates the importance of collaborative information. However, it cannot capture the complex sequential transitions within the session.

Compared with traditional methods, neural network based methods usually have better performance for session-based recommendation. 
In sprite of preforming worse than Item-KNN on Diginetica, GRU4Rec, as the first RNN based method for SBR, still demonstrates the capability of RNN in modeling sequences.
Nevertheless, it relies entirely on RNN to model the complex transitions within the session, which can only captures the point-wise dependencies and may generate fake dependencies \cite{ijcai2019-883}.
%

The subsequent
methods, NARM and STAMP outperform GRU4REC
significantly. NARM combines RNN and attention mechanism, which uses the last hidden state of RNN as the main preference of user, this result indicates that directly using RNN to encode the session sequence may not be sufficient for SBR as RNN only models one way item-transition between adjacent items in a session. We also observe that STAMP, a complete attention-based method, achieves better performance than NARM on Tmall, which incorporates a self-attention over the last item of a session to model the short-term interest, this result demonstrates the effectiveness of assigning different attention weights on different items for session encoding. Compared with RNN, attention mechanism appears to be a better option, although STAMP neglects the chronological order of items in a session.

CSRM and CoSAN performs better than NARM and STAMP over three datasets, which shows the effectiveness of using item transitions from other sessions.
However, the memory networks used by CSRM have limited slots and both of them treat other sessions as a whole one
without distinguishing the relevant item-transitions from the irrelevant ones encoded in other sessions.

By modeling every session sequence as a subgraph and applying GNN to encode items, SR-GNN and GCE-GNN demonstrate the effectiveness of applying GNN in session-based recommendation.
This indicates that the graph modeling would be more suitable than the sequence modeling, RNN, or a set modeling, the attention modeling for SBR.
Our approach SRGI-FM and SRGI-CM outperforms SR-GNN and GCE-GNN on all the three datasets. Specifically, SRGI-CM outperforms the GCE-GNN by $ 1.5 \% $ on Diginetica, $ 13.6 \%$ on Tmall and $ 4.4 \%$ on Nowplaying on average.
Different from SR-GNN and CoSAN, our approach integrates item-level transition information from global context, i.e., other session, and local context, i.e., the current session, and also incorporates reversed position information, leading to consistent better performance.

\subsection{Impact of Global Feature Encoder (RQ2)}
In this section, we aim to study the effect of global transition information and the impact of different graph parameters (\ie graph depth, the number of neighbors and trade-off parameter $\lambda_C$) by conducting experiments over three datasets.

\subsubsection{Effect of global transition information.}
From Table \ref{tab:results}, we can observe that both SRGI-FM and SRGI-CM achieve better performance than B-GNN, which verifies that global transition information can provide useful information for learning the preference of the current session. Comparing with two version of SRGI, SRGI-CM performs better than SRGI-FM in most instances. It is because SRGI-FM directly incorporates the neighbors' features from the global graph, which easily introduces extra noise into the current session. Although SRGI-FM leverages session aware attention mechanism to reduce the influence of noise, the performance of the model is still affected. In comparison, SRGI-CM is less influenced by the noise information as it does not need to fuse global features directly into the current session representation. Specifically, SRGI-CM obtains the representation of each session based on the item features within the session and utilizes contrastive learning to preserve global proximity. The global proximity forces the neighboring items more proximity while different items to be discriminating in the learned embedding space, which benefits the item representation learning and the prediction of the current session. The results in Table \ref{tab:results} demonstrate the effectiveness of our proposed two versions of SRGI.

\begin{figure*}[t]
    \centering
    \subfloat[MRR@20 on Diginetica.]{
    \centering
    \includegraphics[width=0.33\linewidth]{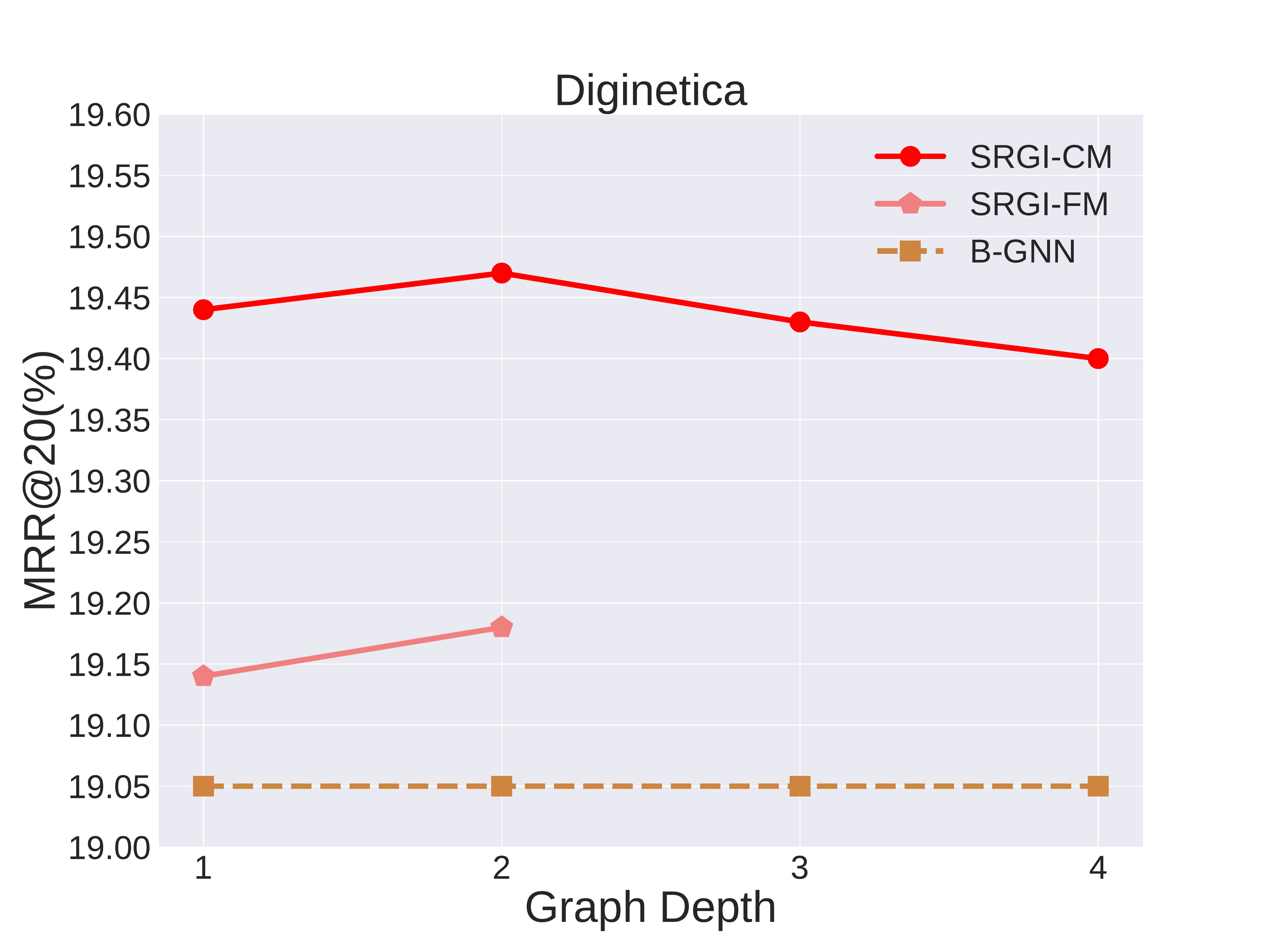}
    \label{fig:DepthMrrDiginetica}
    }
    \subfloat[MRR@20 on Tmall.]{
    \centering
    \includegraphics[width=0.33\linewidth]{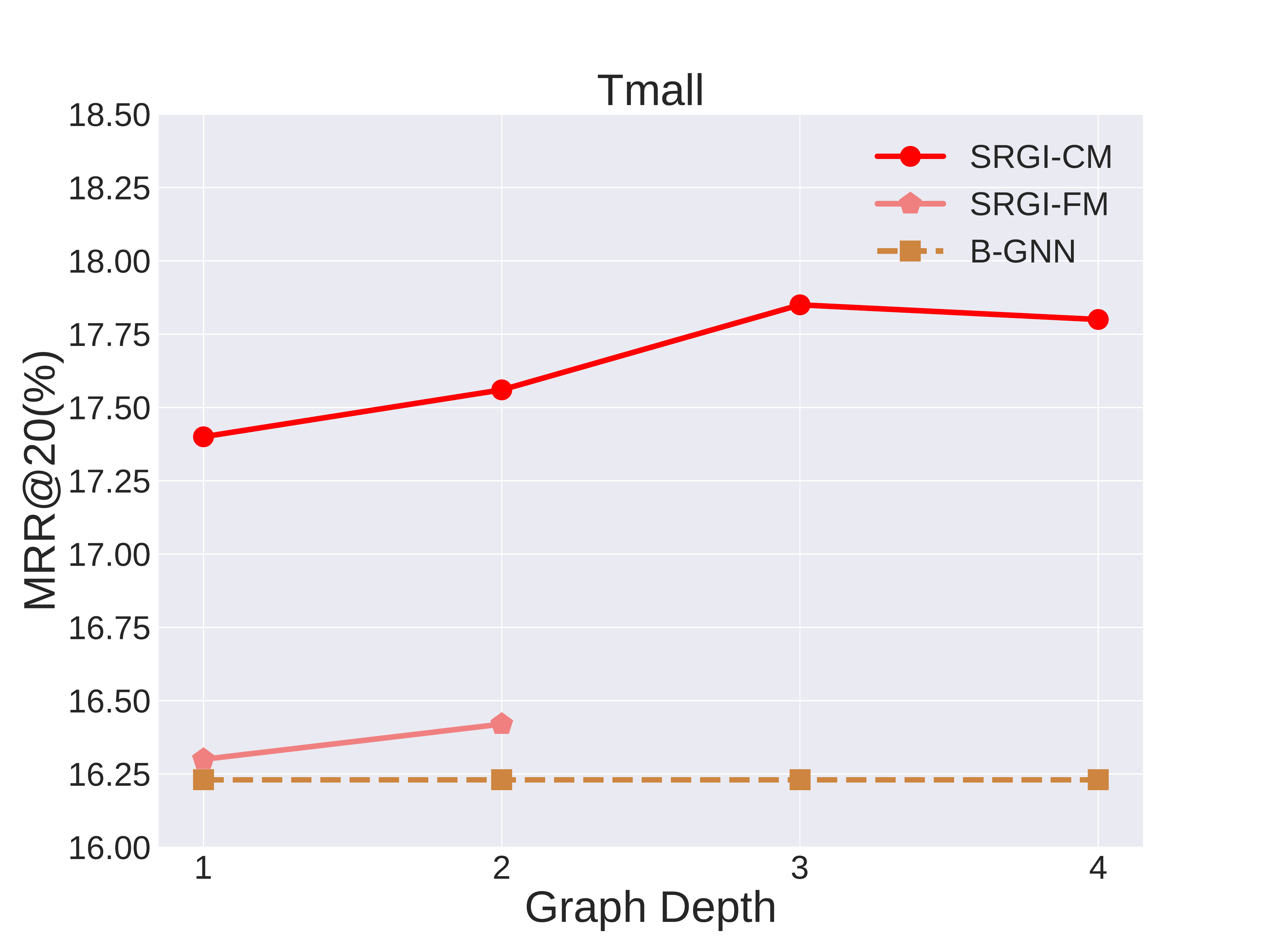}
    \label{fig:DepthMrrTmall}
    }
    \subfloat[MRR@20 on Nowplaying.]{
    \centering
    \includegraphics[width=0.33\linewidth]{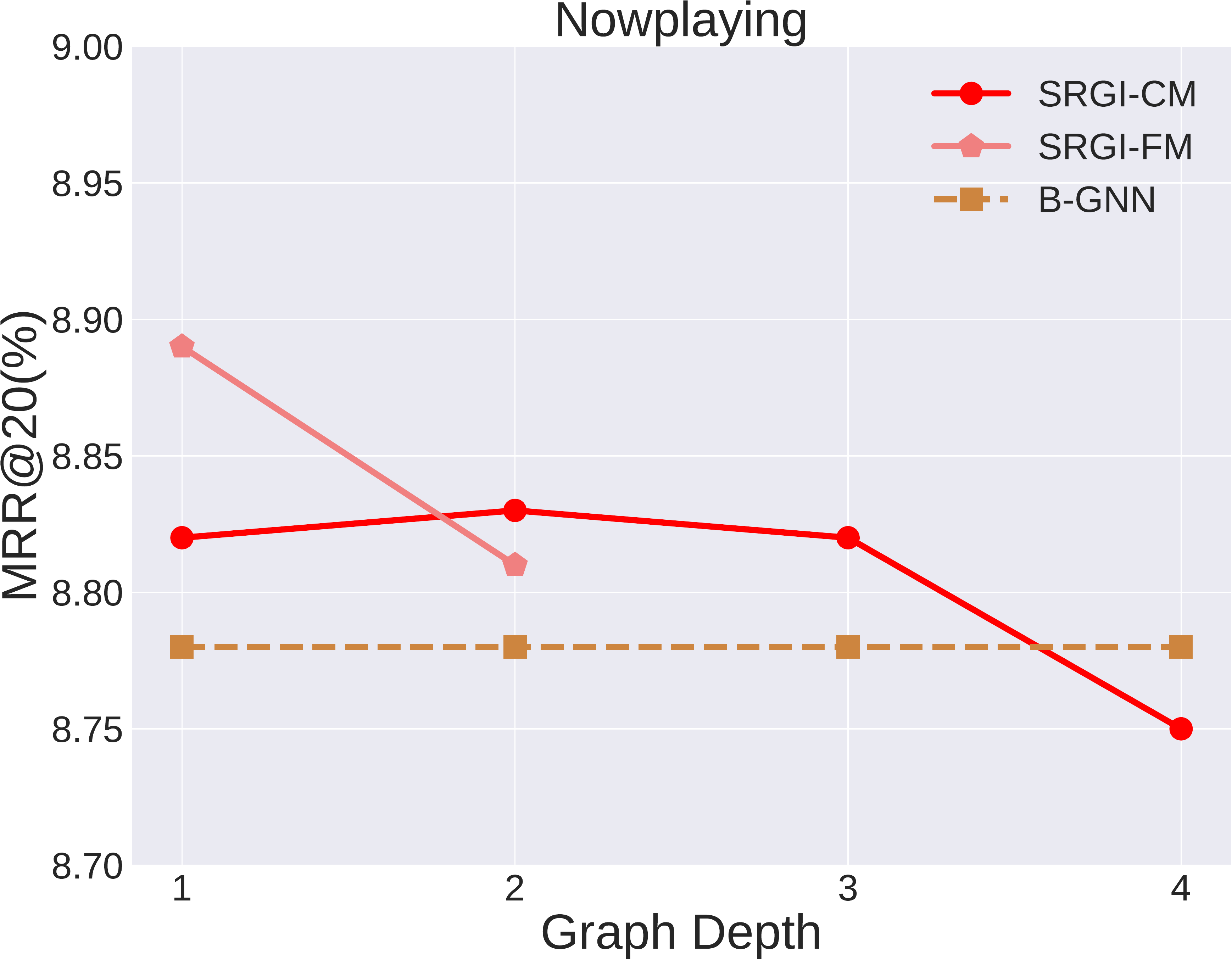}
    \label{fig:DepthMrrNowplaying}
    }

    \subfloat[P@20 on Diginetica.]{
    \centering
    \includegraphics[width=0.33\linewidth]{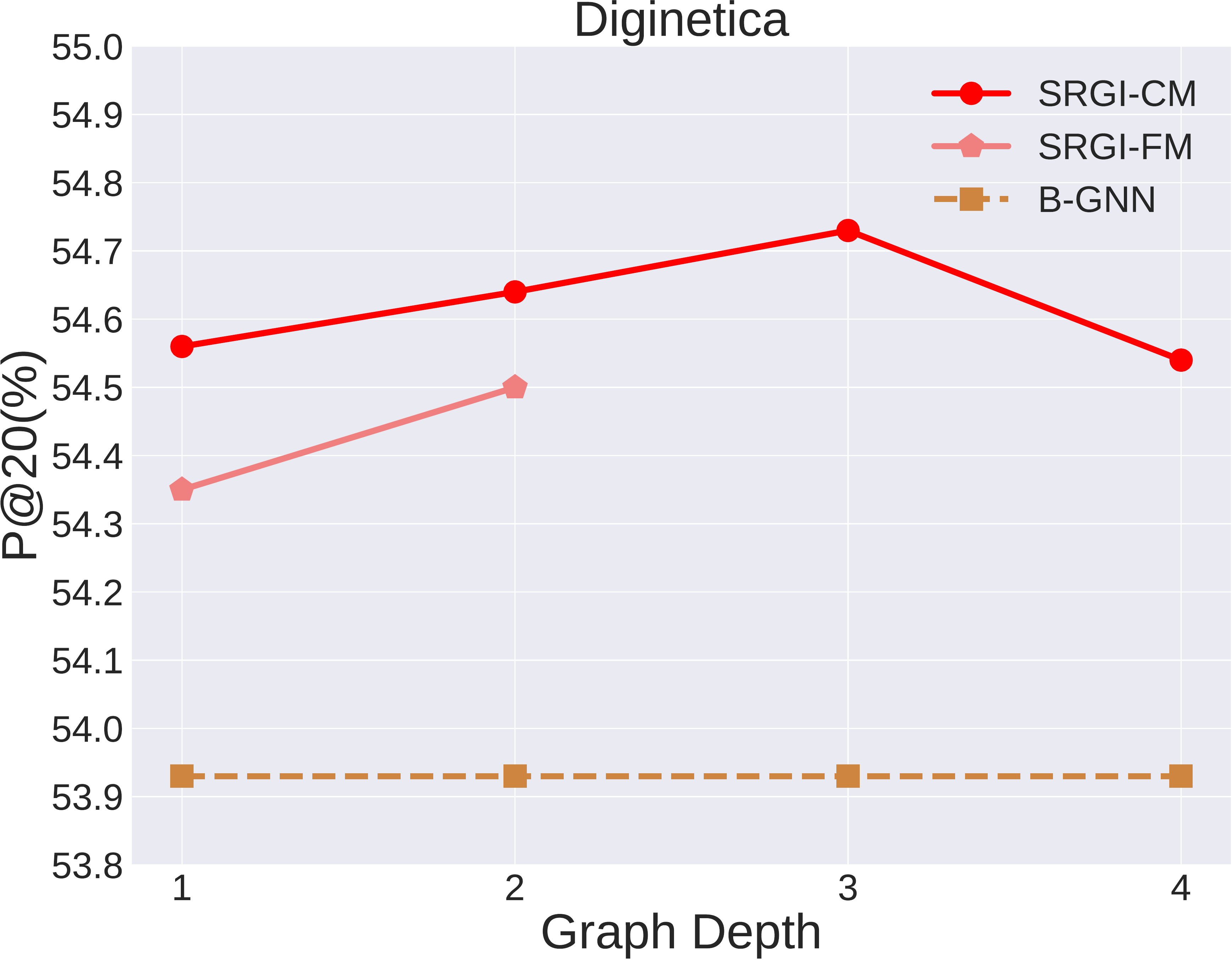}
    \label{fig:DepthPDiginetica}
    }
    \subfloat[P@20 on Tmall.]{
    \centering
    \includegraphics[width=0.33\linewidth]{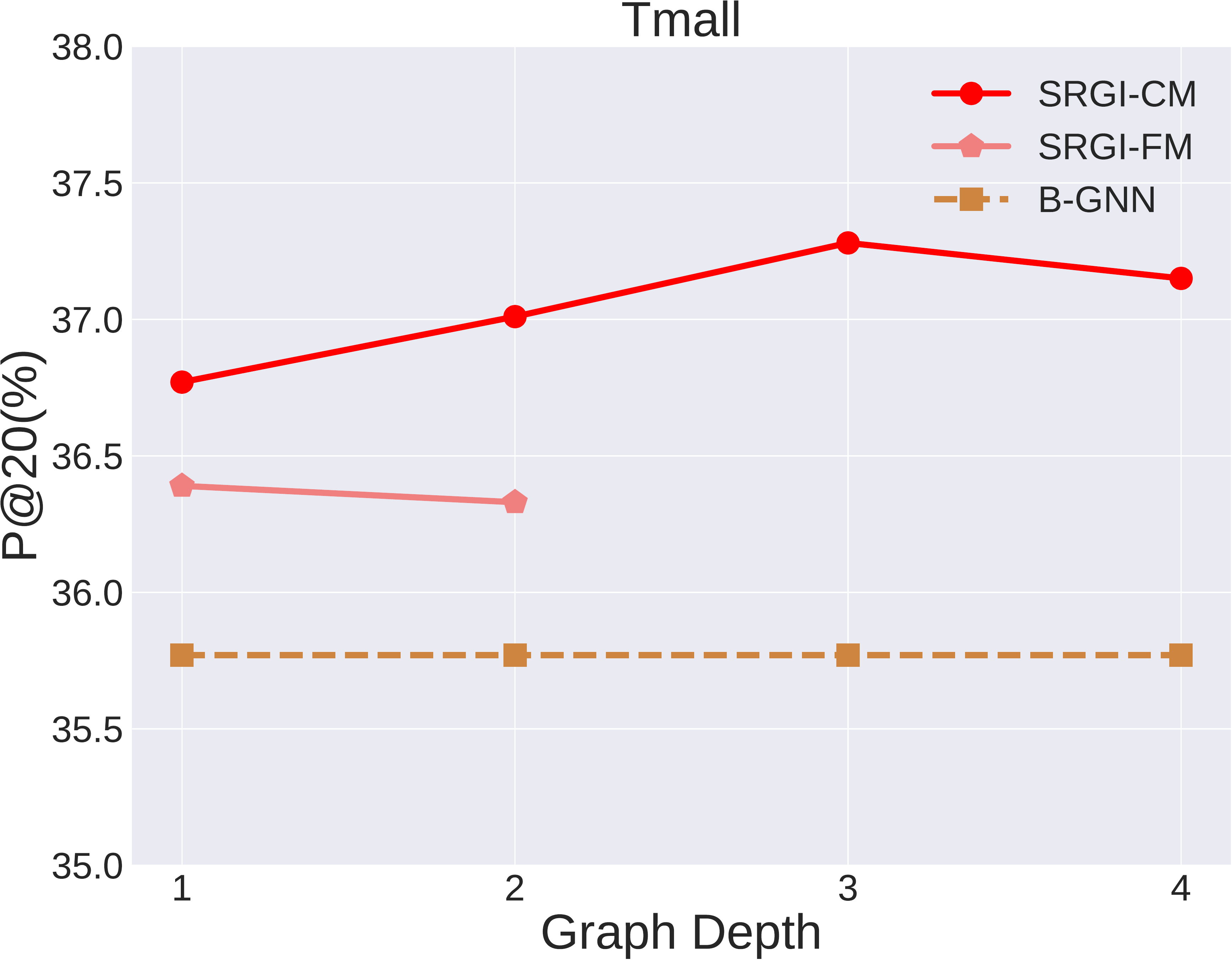}
    \label{fig:DepthPTmall}
    }
    \subfloat[P@20 on Nowplaying.]{
    \centering
    \includegraphics[width=0.33\linewidth]{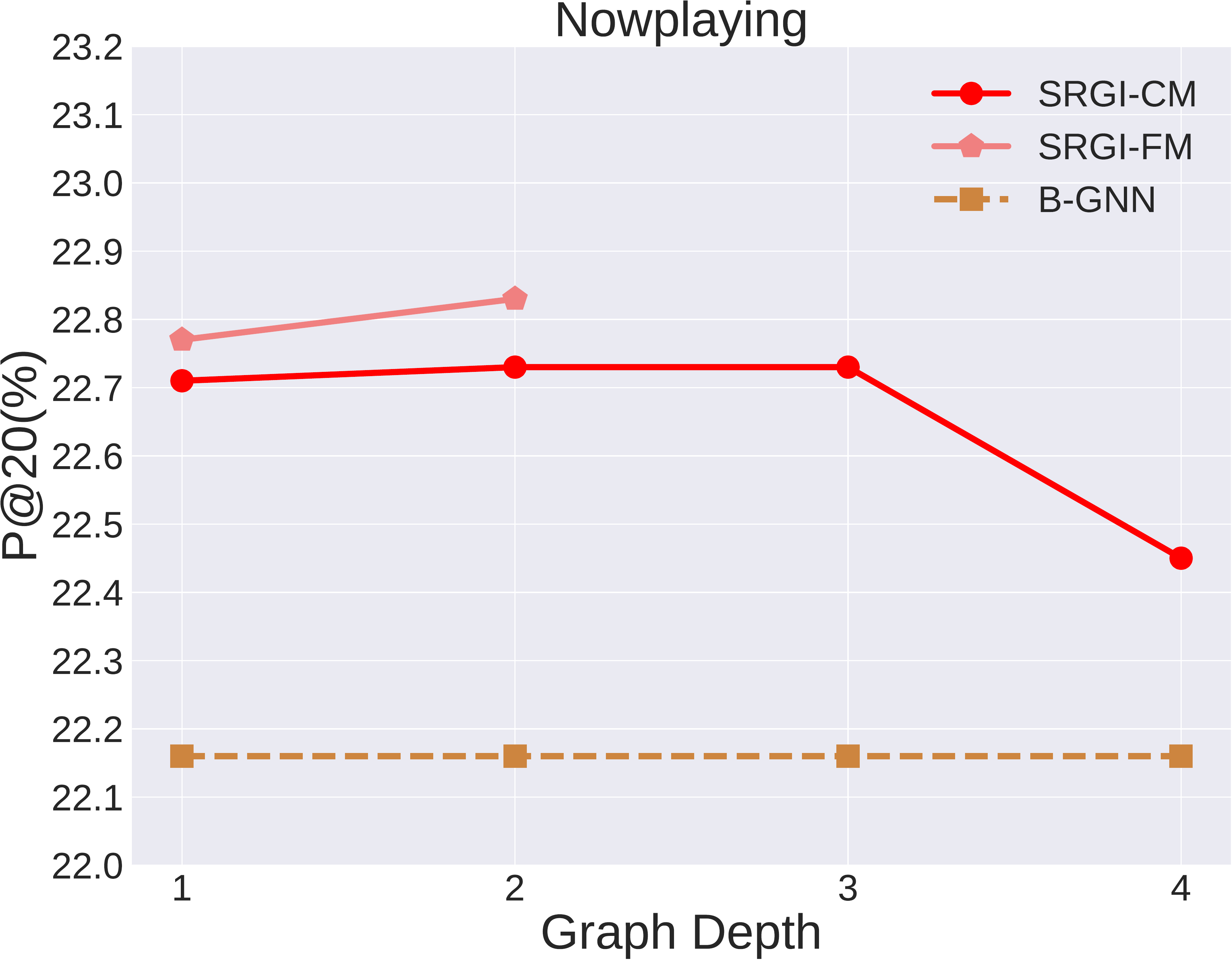}
    \label{fig:DepthPNowplaying}
    }
    \caption{The performance of models with different graph depths.}
    \label{fig:ResultGraphDepth}
\end{figure*}

\subsubsection{Impact of the graph depth.}
We next conduct experiments on three datasets to evaluate the impact of the graph depth on two versions of SRGI. For SRGI-FM, the computational cost is high as it uses attention network in the graph encoder and the number of neighbors of each item increases exponentially as the graph depth increases, therefore we only evaluate the performance of SRGI-FM with 1-hop and 2-hop neighbors due to the limited graphics memory. In contrast, in SRGI-CM we employ GCN architecture, whose computational cost is relatively lower and thus we evaluate the performance of SRGI-CM with the graph depth from 1-hop to 4-hop.

Figure \ref{fig:ResultGraphDepth} shows the performance of two versions of SRGI with different graph depth. It can be observed that both SRGI-CM and SRGI-FM outperform the B-GNN with graph depth over three datasets. Specifically, SRGI-FM with $2$-hop performs better than SRGI-FM with $1$-hop in most situation, which indicates that high-level exploring might obtain more effective information from global graph. Besides, the performance of SRGI-CM is better than SRGI-FM on Diginetica and Tmall, which demonstrates the effectiveness of global proximity for session-based recommendation. Further, we observe the performance of SRGI-CM drops when graph depth is set to 4 on three datasets in terms of P@20 and MRR@20, which shows that higher-level exploring might also introduce more noise during the global proximity learning. Overall, the results in Figure \ref{fig:ResultGraphDepth} demonstrates that the structural information in global graph contain useful global transitions information for current session.

\begin{figure*}[t]
    \centering
    \subfloat[MRR@20 on Diginetica.]{
    \centering
    \includegraphics[width=0.33\linewidth]{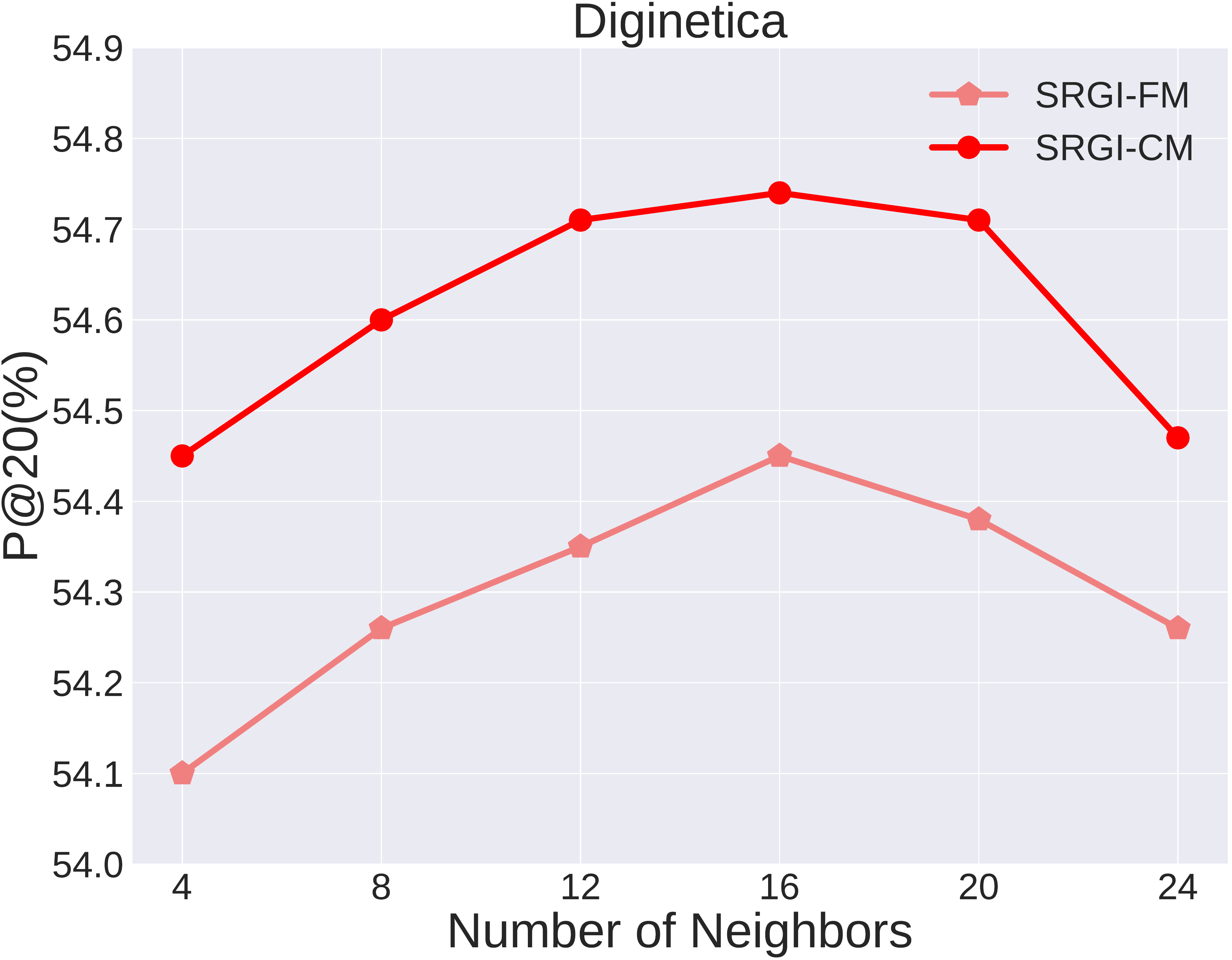}
    \label{fig:NeighborDiginetica}
    }
    \subfloat[MRR@20 on Tmall.]{
    \centering
    \includegraphics[width=0.33\linewidth]{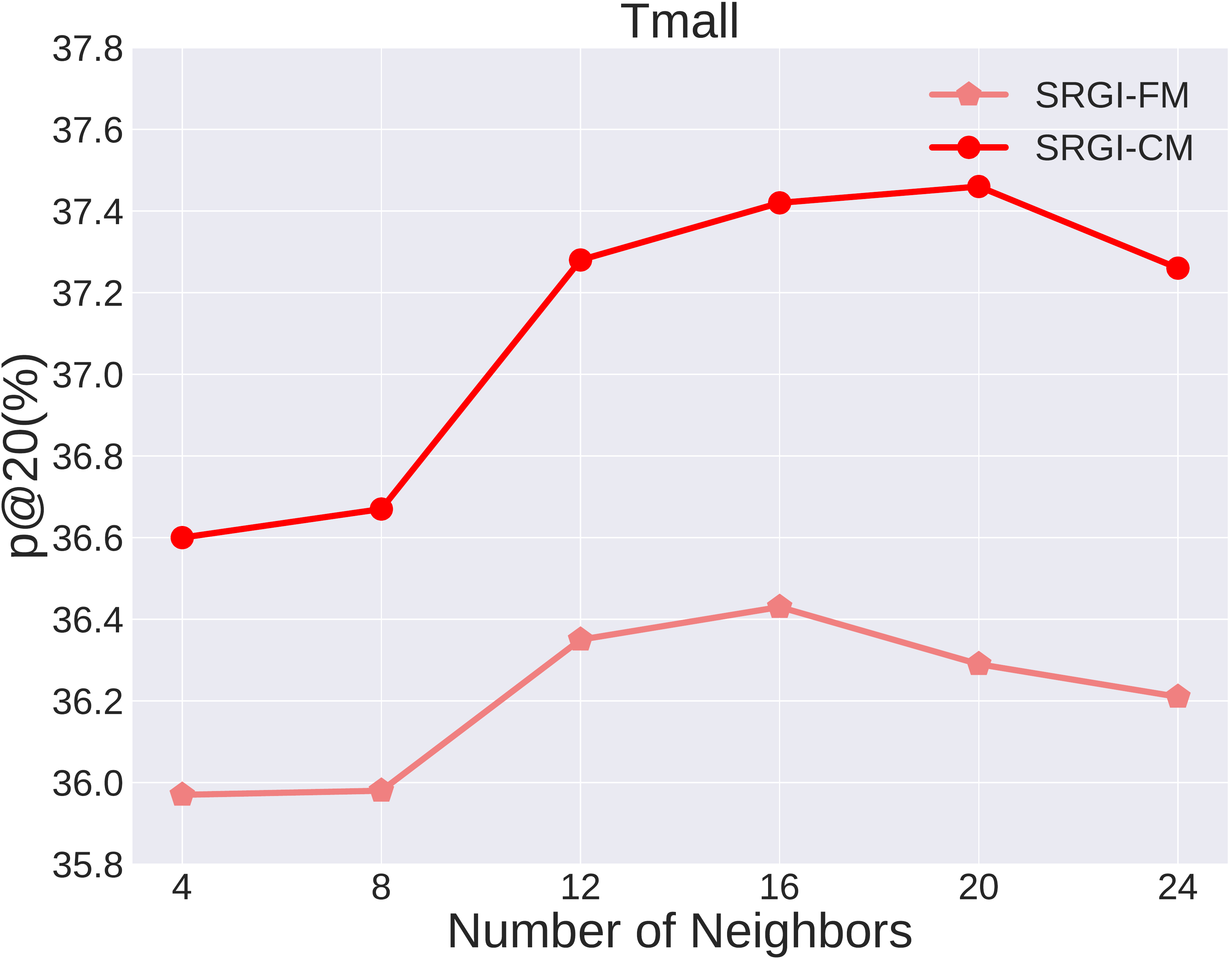}
    \label{fig:NeighborTmall}
    }
    \subfloat[MRR@20 on Nowplaying.]{
    \centering
    \includegraphics[width=0.33\linewidth]{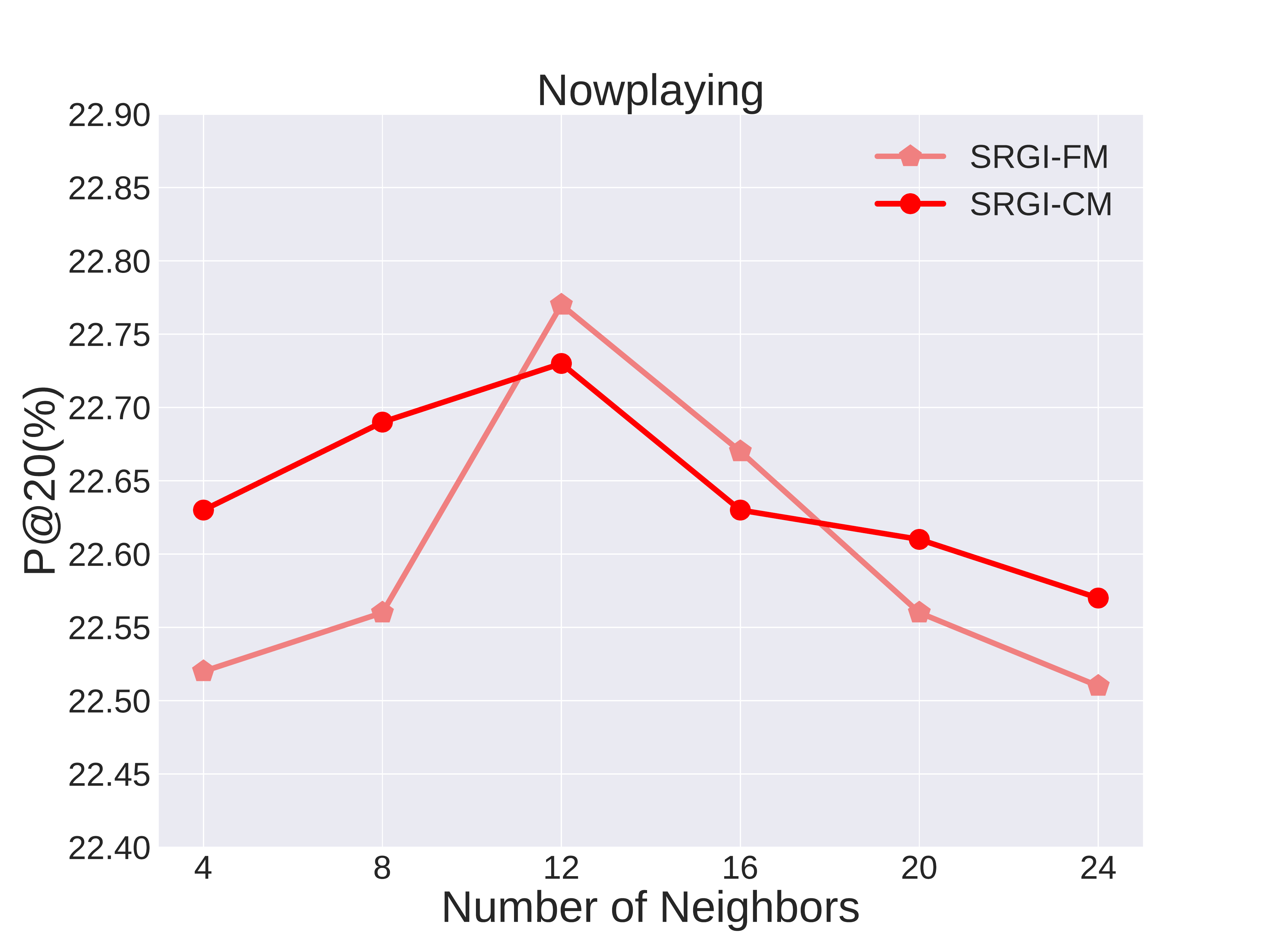}
    \label{fig:Neighborplaying}
    }
    \caption{The performance of models with different number of neighbors.}
    \label{fig:ResultGraphNeighbor}
\end{figure*}

\subsubsection{Impact of the number of neighborhoods on global graph.}
It can be observed from Figure \ref{fig:frameWorkexample} that there will be irrelevant items in a session when user clicked items. As we collect the $\epsilon$-Neighbor set of each item to obtain the global item transitions information, the inclusion of noise was unavoidable. In the proposed method, we only keep top-$N$ edges with the highest weights (\ie frequency) to reduce the influence of noise. When the number of neighbors $N$ increases, more useful global transitions information can be captured, and at the same time more noise will be introduced. Here, we conduct experiment over three datasets to evaluate the impact of $N$ on the performance of our proposed method.

From Figure \ref{fig:ResultGraphNeighbor}, we can observe that with the number of neighbors increase from 4 to 12, the performance of SRGI-CM in terms of MRR@20 becomes better over three datasets. It is because SRGI-CM can capture more effective global transitions information from the neighbors of each item, which benefits the prediction of current session. However, the performance of SRGI-CM drops when it has more neighbors in the global graph, which is affected by the increasing noise in the neighbors.


\subsubsection{Impact of the trade-off parameter $\lambda_C$ in SRGI-CM.}
Trade-off parameter $\lambda_C$ is an important scalar, which controls the influence of global proximity. Higher $\lambda_C$ means that SRGI-CM pays more attention to the global-level item transitions. Therefore, we conduct experiments to study the impact of $\lambda_C$ on the performance of SRGI-CM over three datasets.
From Table \ref{tab:lambda} we observe that with the $\lambda_C$ increases, the performance of SRGI-CM improves in terms of P@20 on Diginetica and Nowplaying, which shows the effective of global information to the current session. However, the performance of SRGI-CM drops on Nowplaying in terms of MRR@20 when the $\lambda_C$ increases. This is because too much attention to global information will affect the learning of the current session. We consider that it is appropriate to set $\lambda_C$ to $50$ or $100$ for most of the scenarios.

{
\setlength{\tabcolsep}{1pt}
\begin{table}[!t]
    \small
	\centering
	\caption{The performance of SRGI-CM with different $\lambda_C$.}
	\label{tab:resultLambda}
	\begin{tabular}{c|cc|cc|cc}
	\hline
		{ \text{Dataset} } & \multicolumn{2}{c|}{\text{Diginetica}} &  \multicolumn{2}{c|}{\text{Tmall}} &  \multicolumn{2}{c}{\text{Nowplaying}}\\
		\hline
		{ \text{Measures} }  & P@20 & MRR@20 & P@20 & MRR@20 & P@20 & MRR@20 \\
		\hline
		{ \text{$\lambda_C = 10$} }  & 54.26 & 19.33 & 36.83 & 17.02 & 22.52   & \textbf{8.92} \\
		{ \text{$\lambda_C = 50$} }  & 54.61 & 19.37 & \textbf{37.28} & 17.85 & 22.83 & 8.86 \\
		{ \text{$\lambda_C = 100$}}  & \textbf{54.71} & 19.32 & 36.63 & 17.77 & \textbf{22.86} & 8.73  \\
		{ \text{$\lambda_C = 150$} }  & \textbf{54.71} & \textbf{19.43} & 35.84 & \textbf{18.02} & 22.72 & 8.55 \\
		{ \text{$\lambda_C = 200$}}  & 54.57 & 19.47 & 35.03 & 17.98 & 22.75 & 8.49  \\
	\hline
\end{tabular}
\label{tab:lambda}
\end{table}
}

{
\setlength{\tabcolsep}{1pt}
\begin{table}[!t]
    \small
	\centering
	\caption{The performance of contrast models.}
	\begin{tabular}{c|cc|cc|cc}
	\hline
		{ \text{Dataset} } & \multicolumn{2}{c|}{\text{Diginetica}} &  \multicolumn{2}{c|}{\text{Tmall}} &  \multicolumn{2}{c}{\text{Nowplaying}}\\
		\hline
		{ \text{Measures} }  & P@20 & MRR@20 & P@20 & MRR@20 & P@20 & MRR@20 \\
		\hline
		{ \text{B-GGNN} }  & 53.89 & 18.81 & 34.88 & 14.89 & \textbf{22.35} & 7.80 \\
		{ \text{B-GAT} }  & 53.61 & 18.85 & 35.16 & 15.40 & 21.70 & 8.20 \\
		{ \text{B-GNN} }  & \textbf{53.93} & \textbf{19.05} & \textbf{35.77} & \textbf{16.23} & 22.16 & \textbf{8.78}  \\
	\hline
\end{tabular}
\label{table:gnns}
\end{table}
}

{
\setlength{\tabcolsep}{1pt}
\begin{table}[!t]
    \small
	\centering
	\caption{The performance of contrast models.}
	\begin{tabular}{c|cc|cc|cc}
	\hline
		{ \text{Dataset} } & \multicolumn{2}{c|}{\text{Diginetica}} &  \multicolumn{2}{c|}{\text{Tmall}} &  \multicolumn{2}{c}{\text{Nowplaying}}\\
		\hline
		{ \text{Measures} }  & P@20 & MRR@20 & P@20 & MRR@20 & P@20 & MRR@20 \\
		\hline
		{ \text{B-GNN-WP} }  & 53.08 & 18.54 & 34.66 & 16.04 & 21.43 & 8.08 \\
		{ \text{B-GNN-FP} }  & 53.64 & 18.76 & 35.36 & 16.19 & 22.07 & 8.65 \\
		{ \text{B-GNN} }  & \textbf{53.93} & \textbf{19.05} & \textbf{35.77} & \textbf{16.23} & \textbf{22.16} & \textbf{8.78}  \\
	\hline
\end{tabular}
\label{table: position}
\end{table}
}

\subsection{Comparison with Different Session-level item representation learning methods. (RQ3)}
In this work, we use mean pooling-based GNNs to convey information from item's neighbors to item itself in the session graph. Specifically, We consider three different kinds of relations of each item and utilize fully connected layer to obtain the new representation of each item. As the representation learning of items is significant for session based recommendation, we conduct experiments to compare our proposed B-GNN with a series of contrast models:
\begin{itemize}
\item B-GGNN: B-GNN with gated GNNs replacing the mean pooling-based GNNs.
\item B-GAT: B-GNN with GAT replacing the mean pooling-based GNNs.
\end{itemize}
Table \ref{table:gnns} shows the performance of B-GNN and different contrast models. We can observe that the proposed B-GNN outperforms B-GGNN and B-GNN-GAT on Diginetica and Tmall. Gated GNNs introduce gated recurrent units into GNNs and GAT leverages attention mechanism into graph, which enhance the ability of graph neural network to extract features. However, more parameters and nonlinear layers will make the network more prone to overfitting and these models do not consider three relations in the session graph. Our GNNs layer utilize mean pooling layer to reduce the parameters and capture three kinds of relations in the session graph.
The results in Table \ref{table:gnns} demonstrate the effectiveness of our proposed GNNs layer.

\subsection{Comparison with Different Session Encoder. (RQ4)}
For efficiently learning the importance of each item in the session sequence, we propose reversed-position aware session encoder. It incorporates the reversed position information with session information, which is used to drive our model to learn the contribution of each item in the current session. To verify this and evaluate the effectiveness of using the position vector in a reverse order, which is proposed in our method, we design a series of contrast models:
\begin{itemize}
\item B-GNN-WP: B-GNN without using position information when aggregating the item features.
\item B-GNN-FP: B-GNN with forward position vector replacing the reversed position vector.
\end{itemize}
Table \ref{table: position} shows the performance of different contrast models.
We can observe that
B-GNN-WP's performance is not satisfactory, as the model is unable to capture the chronological information in the sessions without position vector.
B-GNN-FP performs better than B-GNN-WP over three datasets, which demonstrates the importance of position information. However, the forward position vector can not learn the distance between each item and the target item, which makes the improvement limited.
Our attention network with reversed position vector performs better than the B-GNN-FP over three datasets, it is because reversed position vector contains the relative position information between the current item and the target item. The results demonstrate the effectiveness of reversed position vector and superiority of our session encoder. 

\section{Conclusion}
This paper studies the problem of session-based recommendation, which is a challenging task as
the user identities and historical interactions are often unavailable in real-world scenarios.
It presents two different ways to leverage the high-order global information for session-based recommendation via GNN:
(i) SRGI-FM, which recursively incorporates the neighbors’ feature of each node on global graph via session-aware attention mechanism;
and (ii) SRGI-CM, which treats session-based recommendation as a multi-task learning problem and utilizes graph contrastive learning for preserving global proximity
to learn item representations,
Furthermore, it capture three kinds of relations within the session and incorporates the reversed position embedding to better learn the contribution of each item. 
Comprehensive experiments demonstrate that the proposed method significantly outperforms state-of-the-art baselines over three benchmark datasets consistently, indicating 
it can be effectively used to solve real-world session-based recommendation problems.



\section*{Acknowledgments}
This work was supported in part by the National Natural Science Foundation of China under Grant No.61602197 and Grant No.61772076, and in part by Equipment Pre-Research Fund for The 13th Five-year Plan under Grant No.41412050801.

\ifCLASSOPTIONcaptionsoff
  \newpage
\fi

\bibliographystyle{IEEEtran}
\bibliography{reference}

\end{document}